\documentclass[iop,apj]{emulateapj}
\usepackage{xspace}
\usepackage{amsmath}
\graphicspath{{figures/}}

\usepackage{enumitem}
\newcommand{\phx}{\textsc{Phoenix}\xspace}
\newcommand{\zsun}{$Z_\odot$\xspace}

\newcommand{\snia}{SN~Ia\xspace}
\newcommand{\sneia}{SNe~Ia\xspace}
\newcommand{\nifs}{$^{56}$Ni\xspace}
\newcommand{\caft}{$^{40}$Ca\xspace}
\newcommand{\feff}{$^{54}$Fe\xspace}
\newcommand{\site}{$^{28}$Si\xspace}

\shorttitle{Progenitor Metallicity from SN~Ia Spectra}

\begin{document}
\submitted{Submitted to the Astrophysical Journal on August 24, 2015. Accepted on April 9, 2016}
\title{On Measuring the Metallicity of a Type Ia Supernova's Progenitor}
\author{%
Broxton~J. Miles\altaffilmark{1},
Daniel~R. van Rossum\altaffilmark{2},
Dean~M. Townsley\altaffilmark{1},
F.~X. Timmes\altaffilmark{6,7}
Aaron~P. Jackson\altaffilmark{3},
Alan~C. Calder\altaffilmark{3,4},
and
Edward~F. Brown\altaffilmark{5,6}
}

\altaffiltext{1}{%
Department of Physics \& Astronomy,
University of Alabama, Tuscaloosa, AL; Dean.M.Townsley@ua.edu
}
\altaffiltext{2}{%
Department of Astrophysics,
The University of Chicago, Chicago, IL
}
\altaffiltext{3}{%
Department of Physics \& Astronomy,
The State University of New York - Stony Brook, Stony Brook, NY
}
\altaffiltext{4}{%
New York Center for Computational Sciences,
The State University of New York - Stony Brook, Stony Brook, NY
}
\altaffiltext{5}{%
Department of Physics \& Astronomy,
Michigan State University, East Lansing, MI
}
\altaffiltext{6}{%
The Joint Institute for Nuclear Astrophysics
}
\altaffiltext{7}{%
School of Earth and Space Exploration,
Arizona State University, Tempe, AZ
}

\begin{abstract}

In Type Ia Supernovae (\sneia), the relative abundances of chemical elements are affected by the neutron excess in the composition of the progenitor white dwarf.
Since these products leave signatures in the spectra near maximum light, spectral features may be used to constrain the composition of the progenitor.
We calculate the nucleosynthetic yields for three \snia simulations, assuming single degenerate, Chandrasekhar mass progenitors, for a wide range of progenitor metallicities, and calculate synthetic light curves and spectra to explore correlations between progenitor metallicity and the strength of spectral features.
We use two 2D simulations of the deflagration-detonation-transition scenario with different $^{56}$Ni yields and the W7 simulation to control for differences between explosion models and total yields.
While the overall yields of intermediate mass elements (16 $<$ A $\leq$ 40) differ between the three cases, trends in the yields are similar.
With increasing metallicity, $^{28}$Si yields remain nearly constant, $^{40}$Ca yields decline, and Ti and $^{54}$Fe yields increase.
In the synthetic spectra, we identify two features at 30 days post explosion that appear to deepen with progenitor metallicity: a Ti feature around 4200\,\AA\ and a Fe feature around 5200\,\AA\@.
In all three simulations, their pseudo equivalent widths show a systematic trend with progenitor metallicity.
This suggests that these two features may allow differentiation among progenitor metallicities of observed \sneia and potentially help reduce the intrinsic Hubble scatter.

\end{abstract}


\section{Introduction}
Type Ia supernovae are generally considered to be the result of the thermonuclear disruption of carbon-oxygen white dwarfs in binary star systems.
Although the path which leads to these explosions is uncertain, the results are easily seen and quite useful.
Powered by the radioactive decay of $^{56}$Ni \citep{colgate_69}, \sneia have luminosities on the order of $\sim4\times10^{9}$ L$_\odot$.
This extreme luminosity in conjunction with a generally well followed correlation between peak brightness and decline rate \citep{phillips:absolute,Phillips_99} allow \sneia to be used as standard candles out to redshifts of z$\approx$1.
Measurement of the expansion history of the universe in this redshift range demonstrated that the expansion of the universe is accelerating due to the influence of dark energy \citep{riess.filippenko.ea:observational,perlmutter.aldering.ea:measurements}.

Though the Type Ia class as a whole is generally homogeneous, variations between individual events' spectra and light curves do occur\citep{Branch_1993,Branch_2006,Blondin_2012}. 
Many of these variations can be traced back to the amount of iron-group material produced in the explosion with a portion of this material being radioactive \nifs .
Explosions with lower amounts of iron-group material, and thus lower amounts of \nifs , are redder, dimmer, and have faster decline times than those with higher amounts of iron-group material.
One factor that can affect the production of stable and radioactive iron-group material is the metallicity of the progenitor.
Higher metallicity progenitors have larger abundances of neutrons causing greater amounts of stable iron-group material to be produced and less radioactive Ni \citep{timmes_2003_aa}.

Studies of explosion models used in this work (described below) confirm this result holds when considering the effect of the metallicity on the flame speed in addition to composition \citep{Townsleyetal09,Jackson10}.
The capability to observationally constrain the metallicity of \snia\ progenitors would reduce some of the uncertainty caused by intrinsic variations between individual events.
In this study, we seek possible spectral features that could serve as indicators of progenitor metallicity.

Observationally, it has been clear for some time that environment plays a role in the observed behavior of \sneia.
\cite{Hamuy_2000} saw in a sample of 62 \sneia that brighter, slower declining events preferentially occurred in spiral galaxies, while elliptical galaxies were host primarily to quickly declining, dim events.
Also, they find that metal-poor environments produced brighter \sneia.
\cite{Kelly_2010} find that \sneia that occur in larger, more massive galaxies were 10\% brighter than other \sneia with similar light curve shapes.
They claim that this trend could be caused by a correlation between galaxy mass and metallicity, older progenitors in higher mass galaxies, or other factors such as dust.
Additionally, \cite{Ellis_2008} found that for 36 intermediate-redshift \sneia the differences between observed UV spectra could not be accounted for by dust alone requiring some other environmental factors such as metallicity to play a role.
And, \cite{Howell_2009} found that \sneia that occured in host galaxies with higher metallicities produced up to 10\% less $^{56}$Ni. 
These observational studies all use host galaxy metallicity as a proxy for the progenitor metallicity.
However as metallicity can vary by location and time in galaxies, it would be more effective to have an indicator of the individual progenitor's metallicity in the \snia spectrum itself.

There have been a number of investigations into the relationship between spectra and brightness \citep[e.g.][]{Nugentetal1995,Baileyetal2009,Blondinetal2011,BlondinMandelKirshner2011,Blondin_2012},
mostly with the broad aim of improving the calibration of SN absolute brightness and thereby decreasing the scatter from the Hubble law.
In this present study, we do not attempt to resolve this broader question, but instead focus on evaluating, from theory, candidate indicators of the metallicity of the progenitor that are not too blended or obscured by neighboring features.
Since metallicity is thought to be mostly a secondary parameter relating to the SN brightness, the next step would be to attempt to separate it from the primary parameter.
Given our limited sampling of theoretical models and their current uncertainty, we refrain from addressing primary variation in brightness directly here, and instead limit our study to searching for candidate spectral indicators of metallicity.

Past theoretical studies have focussed on metallicity effects on synthetic spectra, in particular the UV flux, by making sensible ad hoc modifications to the abundance profiles of explosion models (see \citet{Brown15}, and references therein).
Generally, such modifications are not necessarily self-consistent with different levels of progenitor metallicity.
For example, this type of modification does not account for the effect of progenitor metallicity on the \nifs production, which affects post explosion ejecta temperatures, and therefore the shape and color of synthetic light curves and spectra.
\cite{De14} have shown that the production of intermediate mass elements is also affected by progenitor metallicity in a way that is expected to be fairly model independent.
These changes in the yields of the intermediate mass elements (IMEs, with atomic mass number 16 $<$ A $\leq$ 40) should also have visible effects on the observed spectra.
They find the greatest change to be in the abundance of Ca with the least amount of change in the abundance of Si.
This indicates that the characteristic SiII line at 6150\,\AA\ may remain fairly static with metallicity, but the features from other IMEs should vary in such a way that the progenitor metallicity could possibly be measured.

Here, we use multi-dimensional simulations of \sneia using the deflagration-to-detonation transition paradigm with the FLASH hydrodynamics code in conjunction with particle post-processing and the 1-D radiative transfer code PHOENIX to examine the effects of progenitor metallicity on nucleosynthetic yields, light curves, and spectra.
In section 2, we outline the \snia models we are using as well as a description of the hydrodynamic, particle post-processing, and radiative transfer calculations.
In section 3, we discuss the nucleosynthesis results, synthetic light curves, and spectra.
We also compare our results to \cite{Foley_2013} and \cite{Graham15}, who examined SN2011by and SN2011fe, two spectroscopically normal \sneia which had nearly identical decline times and optical spectra, but interesting differences in the NUV region.
Finally, in section 4, we give our conclusions and discuss the possibility of using spectral features as an indicator of progenitor metallicity.

\section{\snia Explosion Simulations}

We consider two types of explosion simulations, both variations of the delayed-detonation model for \sneia \citep{khokhlov91+dd}.
The main model is a 2-dimensional simulation of a Chandrasekhar-mass WD exploding via the deflagration-detonation transition (DDT) mechanism.
These simulations are identical to those we have performed previously to study systematic variation in this scenario \citep{Krueger10,Jackson10,Krueger12}.
For comparison, we also compute nucleosynthetic yields, spectra, and light-curves for the well-studied, 1-dimensional W7 model \citep{Nomo84}.
We briefly describe our 2D explosion model and post-processing here.

\subsection{Hydrodynamic Simulations}

We chose two realizations of a CO white dwarf with a central density $2\times10^{9}$ g/cm$^{3}$.
\cite{Krueger12} created 5 CO white dwarf progenitors with central densities that ranged from $1\times10^{9}$ to $5\times10^{9}$ g/cm$^{3}$.
For each progenitor, 30 realizations were created by applying initial conditions in the form of random perturbations on the initial burned region.
This was done to better understand potential systematic biases such as how the morphology of the initial conditions might influence the results.
Many of their progenitors produced higher yields of \nifs than is typical in \sneia.
In light of this, we selected realizations R01 and R10 due to their lower estimated \nifs yields, 0.8 M$_{\odot}$ and 0.7 M$_{\odot}$ respectively, which are closer to typical values for \snia \citep{Howeetal09}. 
In this paper we refer to these two simulations as DDT-high and DDT-low.

We recalculated the simulations because the original simulations did not feature tracer particles that are needed for this work (see below).
For this work we used a newer version of the FLASH code, version 4.0.
The simulation results are nearly identical to the those published in \citet{Krueger12}, with some small changes due to changes to the hydrodynamics method in the intervening FLASH releases.
The overall software consists of the publically available version of the FLASH software instrument\footnote{available from http://flash.uchicago.edu} along with added components\footnote{available from http://astronomy.ua.edu/townsley/code\label{note3}} for efficient modeling of carbon-oxygen flames and detonations
developed during our previous work \citep{Calderetal07,Townsley.calder.ea:flame,Townsleyetal09,Jackson10,Krueger12}
as well as software for computing nucleosynthesis in post-processing \citep{Townsley15}$^{\ref{note3}}$.
In addition to the hydrodynamics itself, the current public FLASH software release includes the advection-diffusion treatment used to model the propagation of the deflagration.
The FLASH modules implementing the carbon-oxygen burning model$^{\ref{note3}}$ will also be included in a future public release of the FLASH software instrument.

Our simulations begin at the end of the core convection phase of the runaway just
after the ignition of a propagating thermonuclear flame (see
\citealt{Nonaka12} for a discussion of this ignition process).  We
make the assumption that the flame is ignited in such a way that it spreads
in the center of the star before plumes rise to lower densities.  This is an
analog of the ``multi-point'' ignition scenarios explored by other authors in
3D (e.g.  \citealt{Seitenzahl13} for an example).
\cite{Nonaka12} suggests that multi-point ignitions are likely
disfavored by nature.  Nevertheless, as others have shown previously and we demonstrate again in this work,
this does lead to an explosion that reproduces spectral observations of
normal \sneia fairly well.  Our simulations follow the nuclear and
hydrodynamic evolution of the flame within the star until it reaches a
density of $10^{7.1}$~g~cm$^{-3}$, at which point it is assumed that a
detonation is ignited by a DDT mechanism
\citep{khokhlov91+dd,Poludnenko11}.  This detonation is ignited with
the introduction of a small heated region since all burning fronts, and
therefore any physical DDT processes, are unresolved in our full-star
simulations.  The complete incineration of the star by the detonation then
ensues.  Active burning fronts cease due to the expansion of the star by
about 1.8 seconds and then we follow the expansion of the star to 4 seconds,
at which point it is already close to force-free expansion.

Our model of carbon-oxygen burning used in hydrodynamics is a three-level burning model that is calibrated against steady state deflagrations and detonations.
The resulting temperature, $T$, and density, $\rho$, in the explosion simulation enable post-processing of Lagrangian tracer particles (see next section) to obtain abundances behind the reaction front with an accuracy of 5-10\% \citep{Townsley15}.
The global energy deposition based on the burning model is consistent with the yields obtained from post-processing to within 10\%.

\subsection{Lagrangian Tracer Particles}
During the explosion simulations, the temperature-density histories of a set of 100,000 Lagrangian tracer
particles are recorded.
These tracer particles are inactive, meaning that they do not affect the hydrodynamics but are only advected with the flow, recording the history of local fluid variables.
During the burning phase of the explosion, the histories are recorded every time step, giving the full time resolution data which is needed for accurate post-processing \citep{Townsley15}.
The equal mass particles are distributed randomly.
After the hydrodynamic simulation is complete, we use
the TORCH\footnote{Available at http://cococubed.asu.edu; Our version available
at http://astronomy.ua.edu/townsley/code} software instrument to compute the evolution of a set
of 225 nuclides subject to the conditions recorded for a given fluid element.
This post-processing step allows the computation of detailed nucleosynthetic yields without having to compute the reaction or advection of all of these species in the main hydrodynamic simulation.

The density and temperature histories for W7 were obtained from Friedel Thielemann (priv comm).
These consist of about 500 time steps for 175 zones out to 4.15 seconds.
These histories were treated in the same manner as Lagrangian tracks from our 2D simulations.
Similar yields were used by \citet{De14} to investigate the dependence of Si-group yields on metallicity in comparison to trends expected from quasi-equilibrium calculations.

\subsection{Nucleosynthesis}
The nucleosynthetic yields from the explosion simulations are determined by post-processing the histories of Lagrangian tracer particles.
The reaction network calculation, described in \cite{Townsley15}, determines mass fractions for each particle.
As described in \citet{Townsley15}, for fluid that is processed by the deflagration we reconstruct, in post-processing, a portion of the Lagrangian history in order to obtain a more realistic $\rho(t)$, $T(t)$ near the artificially thickened flame front.
We do not do so for detonations or for the W7 element histories, instead using the $\rho(t)$, $T(t)$ history directly.
After post-processing, radioactive products with half-lives smaller than 5 days, according to beta decay rates in TORCH, have their calculated abundances converted into their decay products.
Three important decay chains of three isotopes are calculated during the radiation transport simulations: $^{56}$Ni, $^{52}$Fe, and $^{48}$Cr.
The decay chain of $^{56}$Ni to $^{56}$Fe is well understood to be critical in the powering of \sneia light curves.
$^{56}$Ni has a half-life of 6.075 days\footnote{\label{note1}http://www.nndc.bnl.gov/chart/} and transitions by beta decay to $^{56}$Co.
$^{56}$Co has a half-life of 77.236 days$^{\ref{note1}}$ and transitions by beta decay to $^{56}$Fe.
The decay of $^{52}$Fe to $^{52}$Cr is complex.
$^{52}$Fe has a half-life of 8.275 hours$^{\ref{note1}}$ and transitions by beta decay to the 2+ excited state of $^{52}$Mn. 
The excited state has a half-life of 21.1 minutes$^{\ref{note1}}$, but has two possible decay branches.
98.25\% will transition by beta decay to $^{52}$Co, and the remaining 1.75\% decays to the ground state of $^{52}$Mn via internal transition.
The ground state has a half-life of 5.591 days$^{\ref{note1}}$ \citep{Dessart_2014, NDS_52}.
However, we do not track this small amount of $^{52}$Mn due to its negligible influence on spectral features.
$^{48}$Cr has a half-life of 21.56 hours$^{\ref{note1}}$ and decays to $^{48}$V.
The $^{48}$V has a much longer half-life at 15.9735 days$^{\ref{note1}}$, therefore its presence and changing abundance is important throughout the radiation transport calculations.

The particles are then mapped onto a 1D spherically symmetric velocity grid with 100 cells to match the mapping of the radiation transport calculations, using the final velocities each particle has at the end of the simulation.
The mass fractions in each cell of the velocity grid follow from the average mass fractions of all particles in that cell.
These mass fractions per cell, together with a spherically symmetrized density profile from the hydro simulation%
\footnote{This is done with Nathan Hearn's QuickFlash analysis tools, using the latest checkpoint from the hydro simulation for which the ejected material has not begun to run off the grid.}
and the geometrical volume of each cell, give the total yields (in gram) per species.

If a velocity grid cell contains many particles we limit the number of particles that are post-processed to 100, selected randomly.
This reduces the amount of time needed for post-processing.
An independently selected set leads to abundances that differ on the order of 10\% in velocity grid cells that contain greater than 100 particles, and total yields that differ by less than 5\%. 

\subsection{Progenitor White Dwarf Abundances}
The progenitor white dwarf is comprised broadly of two regions: a convective core,  the  inner 0.8 M$_\odot$ of the star, and a non-convective outer region.
Our progenitor abundances are constructed to be similar to those in \cite{DomiHoefStra01}, as well as to account for pre-explosion carbon burning \citep{PiroChan08}.

In this work, the progenitor metallicity is varied by changing the initial abundance state of the tracer particles, not by calculating separate hydro simulations for different initial abundances, because the energy release by carbon-oxygen burning is only marginally modified by the metallicity variations.
The hydrodynamic initial model is at a single metallicity of Z/Z$_\odot$ = 1.33.
The three parameters that control the carbon-oxygen burning model used in the hydro stand for the abundances of $^{12}$C, $^{16}$O, and $^{22}$Ne.
We choose the convective core to be made up of 40\% $^{12}$C, 3\% $^{22}$Ne, with the remaining 57\% as $^{16}$O; the envelope is 50\% $^{12}$C, 48\% $^{16}$O, and 2\% $^{22}$Ne.
The $^{22}$Ne is a stand-in for the neutron excess contributed from the metallicity in both the core and the outer layers \citep{timmes_2003_aa}.
In the convective region the additional 1\% of $^{22}$Ne stands in for the neutron excess from the products of pre-detonation simmering \citep{PiroBildsten2008,Chamulak08}.

\begin{figure}
\centerline{\includegraphics[width=.50\textwidth]{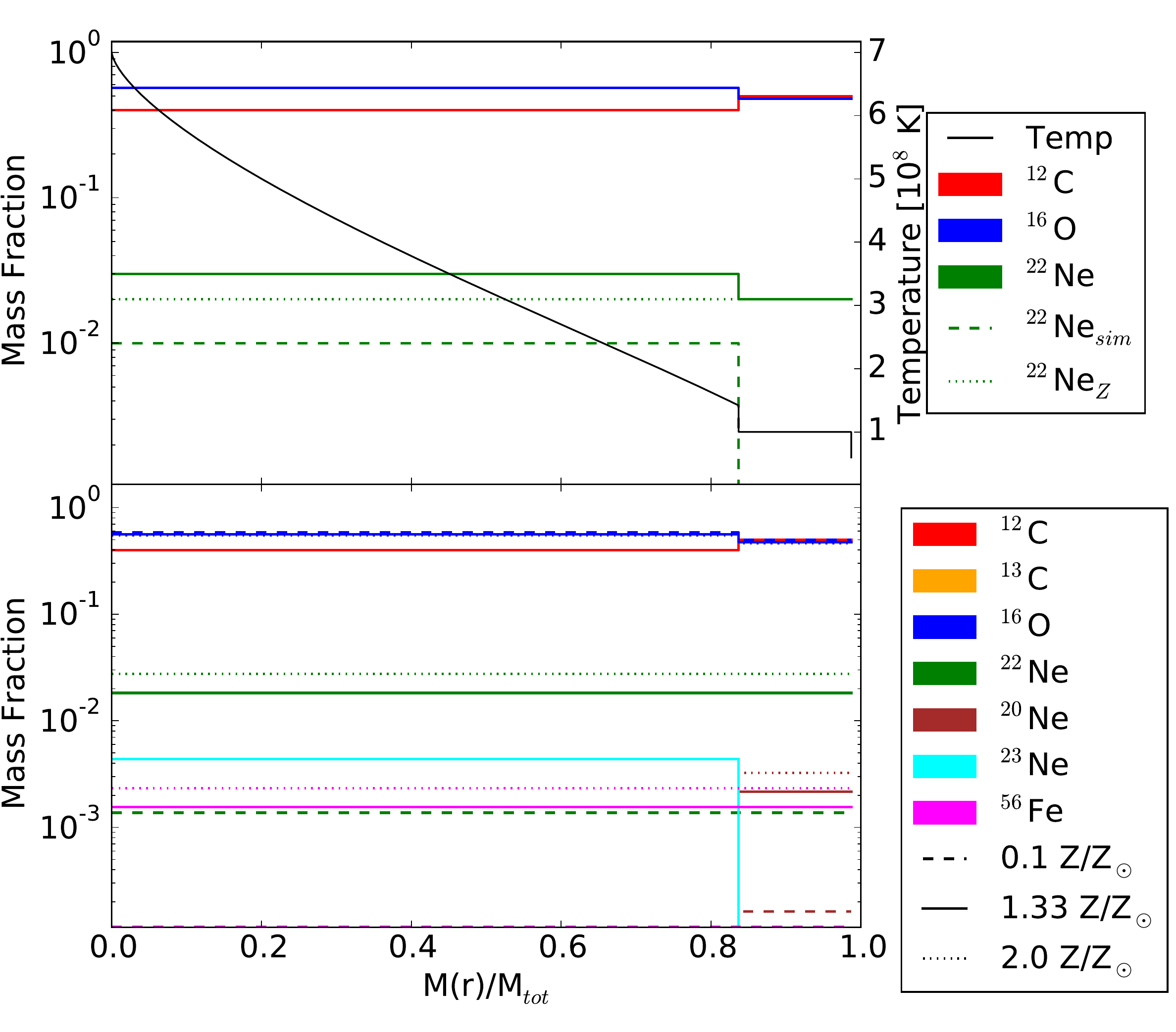}}
\caption{\emph{Top}: The initial composition and temperature (black) profile of the progenitor white dwarf used in the hydrodynamic calculations.
This reduced set of nuclides consists of $^{12}$C (red), $^{16}$O (blue), $^{22}$Ne (green).
$^{22}$Ne is a stand in for the contributions from the simmering ashes (dashed), only present in the convective region, and the metallicity (dash-dot) present throughout the star.
\emph{Bottom:} Initial composition profile used in the post-processing showing a small subset of 225 species used, with values shown for different metallicities.
$^{13}$C (orange), $^{20}$Ne (brown), and $^{23}$Ne (cyan) are all components of the simmering ashes held at equal fixed abundances in the convective region, and consequently, overlap one another.
Outside of the convective region, $^{20}$Ne also contributes as a component of the metals.}\label{fig:model_profile}
\end{figure}

When post-processing the particles with a large nuclear network, more complete and realistic progenitor abundances can be used.
Figure~\ref{fig:model_profile} shows how the material affected by metallicity and the simmering ashes are distributed through the star for both the reduced abundance set and the abundances used for post-processing. 
A particle's initial location is used to determine
whether or not the particle began in the convection region, and thus the initial $^{12}$C fraction and the relative contributions from metallicity and simmering.
The initial abundances for post-processing are then set using 4 components:
\begin{itemize}
\item The mass fraction $X_{^{12}\rm C}$ is set to the value at the particle's initial position in the hydrodynamic progenitor.
\item The abundance of all other nuclides are first set to scaled solar abundances based on $Z/Z_\odot$ and abundances from \citet{anders.grevesse:abundances}.
In doing this, we assume all metals lighter than $^{18}$O, i.e. $^{12}$C, $^{14}$N, and $^{16}$O are converted into $^{22}$Ne and are thus added to that abundance instead.
\item A simmering contribution is added to some abundances for particles that are initially in the convective region.
We approximate that the simmering ashes, of a fraction $X_{\rm sim}$ determined below, are made up of an even mix of $^{13}$C, $^{16}$O, $^{20}$Ne, and $^{23}$Ne \citep{Chamulak08}.
\item Once the $^{12}$C, $^{13}$C, $^{20}$Ne, $^{22}$Ne, $^{23}$Ne and all heavier metal abundances are set in this fashion, the remainder of the material is assumed to be $^{16}$O.
\end{itemize}
For this study, the contribution from simmering ashes is held fixed as the metallicity is varied.  Changing the metallicity thus affects the second and fourth step.

The amount of simmering ashes, $X_{\rm sim}$, is set to obtain consistency with the progenitor metallicity of $Z/Z_\odot=1.33$ used in the hydrodynamics, based on the $Y_e$,
\begin{equation}
Y_{e} = \sum_{i} X_{i}\frac{Z_{i}}{A_{i}},
\end{equation}
in the outer layers.
Here $Z_i$ and $A_i$ are the number of protons and number of nucleons in species $i$.
Matching $Y_e$ between the progenitor and the initial abundances used for post-processing gives
\begin{multline}
\label{eq:yebalance}
X_{^{16}\rm O,prog}\frac{8}{16}+X_{^{12}\rm C,prog}\frac{6}{12}+X_{^{22}\rm Ne,prog}\frac{10}{22} \\
\begin{split}
=&\, X_{^{16}\rm O}\frac{8}{16}+X_{^{12}\rm C}\frac{6}{12}+X_{Z}Y_{e,Z}  \\
 & + X_{\rm sim}\frac14  \left(\frac{8}{16}+\frac{6}{13} +\frac{10}{20}+\frac{10}{23}\right)
\end{split}
\end{multline}
where abundances with the ``prog'' subscript are those found at the initial particle position in the progenitor used in the hydrodynamics,
$X_Z$ is the sum of the scaled solar abundances of all material in the metallicity-scaled component, including $^{22}$Ne, $Y_{e,Z}$ is the $Y_e$ of the metallicity-scaled component, and equal abundances of the 4 simmering
products have been assumed.
For the hydrodynamic abundances given above, $Z/Z_\odot = 1.33 $, solar abundances from \cite{anders.grevesse:abundances}, and $X_{^{16}\rm O}=1-X_{\rm sim}-X_Z-X_{^{12}\rm C}$, Equation \eqref{eq:yebalance} gives $X_{\rm sim} = 0.0175$ in the convection zone and $X_{\rm sim}=0$ outside it.
This value of $X_{\rm sim}$ is then used to construct the initial abundances for post-processing at all metallicities.

\subsection{Radiation Transport Simulations}\label{sec:radtrans}
We calculate light curves and spectra using \phx.
\phx\ is a stellar atmosphere and radiation transport software instrument \citep{Hauschildt99, Hauschildt04, Vanrossum12}.
\phx\ numerically solves the special relativistic radiative transfer equation using the efficient and highly accurate short characteristic and operator splitting methods.
It samples millions of atomic lines individually.
\phx\ solves for the time evolution using the radiation energy balance method.
The code does not use the Sobolev approximation, diffusion approximations or opacity binning approximations.
We operate \phx\ in one dimension, assuming spherical symmetry.

The deterministic radiation transport method employed in \phx\ calculates spectra without the noise that is inherent to Monte Carlo methods.
This allows the study of small effects of individual atomic transition lines on the calculated spectrum through knockout spectra \citep{Vanrossum15}.
Knock-out spectra are calculated by post-processing a normal calculation of light curves and spectra.
For the calculation of knockout spectra, the opacity of individual transition lines is artificially set to zero while keeping the temperatures, electron densities, and atomic occupation numbers fixed to the values determined in the original calculation.
The difference between the original spectrum and a knockout spectrum is attributed to the same line opacity that was artificially set to zero.
Here, as a diagnostic, we use knockout spectra to identify which chemical elements are responsible for the differences that metallicity changes cause in the calculated spectra.

We note that a spectrum is not merely a linear combination of individual lines.
When multiple knockout spectra are shown together in one plot each of the knockout spectra is interpreted as described above, but there is no useful interpretation to a linear combination of knockout spectra.

\section{Results}

\subsection{Nucleosynthetic Yields}
The total yields for six different progenitor metallicities, Z/Z$_{\odot}$ $\in$ [0.1, 0.5, 1.33, 2.0, 3.0, 4.0], of each of the three explosion simulations are listed in Table~\ref{tab:yields_grouped}.
\begin{deluxetable}{l|cccccc}
\tablecaption{Nucleosynthetic yields grouped by atomic mass in $M_\odot$ for three simulations and six progenitor metallicities. \label{tab:yields_grouped}}
\tablehead{%
\colhead{Z/\zsun} & \colhead{0.1} & \colhead{0.5} & \colhead{1.33} & \colhead{2} & \colhead{3} & \colhead{4}
}
\startdata
\phantom{\large X}&\multicolumn{6}{c}{DDT-high}\\
\hline
CO\tablenotemark{a}    & 0.053 & 0.053 & 0.053 & 0.053 & 0.052 & 0.052 \\
IME\tablenotemark{b}   & 0.322 & 0.320 & 0.314 & 0.310 & 0.302 & 0.295 \\
IGE\tablenotemark{c}   & 0.992 & 0.994 & 1.003 & 1.006 & 1.012 & 1.025 \\
\hline
\nifs & 0.848 & 0.831 & 0.800 & 0.775 & 0.738 & 0.703 \\
\hline
\phantom{\Large X}&\multicolumn{6}{c}{DDT-low}\\
\hline
CO    & 0.076 & 0.077 & 0.077 & 0.076 & 0.076 & 0.075 \\
IME   & 0.385 & 0.382 & 0.375 & 0.369 & 0.360 & 0.351 \\
IGE   & 0.906 & 0.909 & 0.916 & 0.922 & 0.932 & 0.941 \\
\hline
\nifs & 0.771 & 0.756 & 0.726 & 0.703 & 0.669 & 0.637 \\
\hline
\phantom{\Large X}&\multicolumn{6}{c}{W7}\\
\hline
CO    & 0.166 & 0.166 & 0.165 & 0.163 & 0.161 & 0.159 \\
IME   & 0.306 & 0.304 & 0.300 & 0.297 & 0.292 & 0.288 \\
IGE   & 0.889 & 0.890 & 0.896 & 0.900 & 0.907 & 0.913 \\
\hline
\nifs & 0.698 & 0.684 & 0.658 & 0.638 & 0.607 & 0.577
\enddata
\tablenotetext{a}{CO: $A = 12$ and $A = 16$}
\tablenotetext{b}{IME: $16< A \leq 40$}
\tablenotetext{c}{IGE: $A>40$}
\end{deluxetable}
The yields are grouped into Carbon and Oxygen (CO, $A = 12$ and $A = 16$), intermediate mass elements (IME, $16< A \leq 40$), and iron-group elements (IGE, $A>40$).
\nifs is also included in the table because of its importance as energy source for light curves and spectra.
Table~\ref{tab:yields_grouped} shows that the amount of radioactive Nickel decreases with increasing progenitor metallicity while the total yields of iron-group elements increases due to the increased neutronization.

\begin{figure}
\centerline{\includegraphics[width=.5\textwidth]{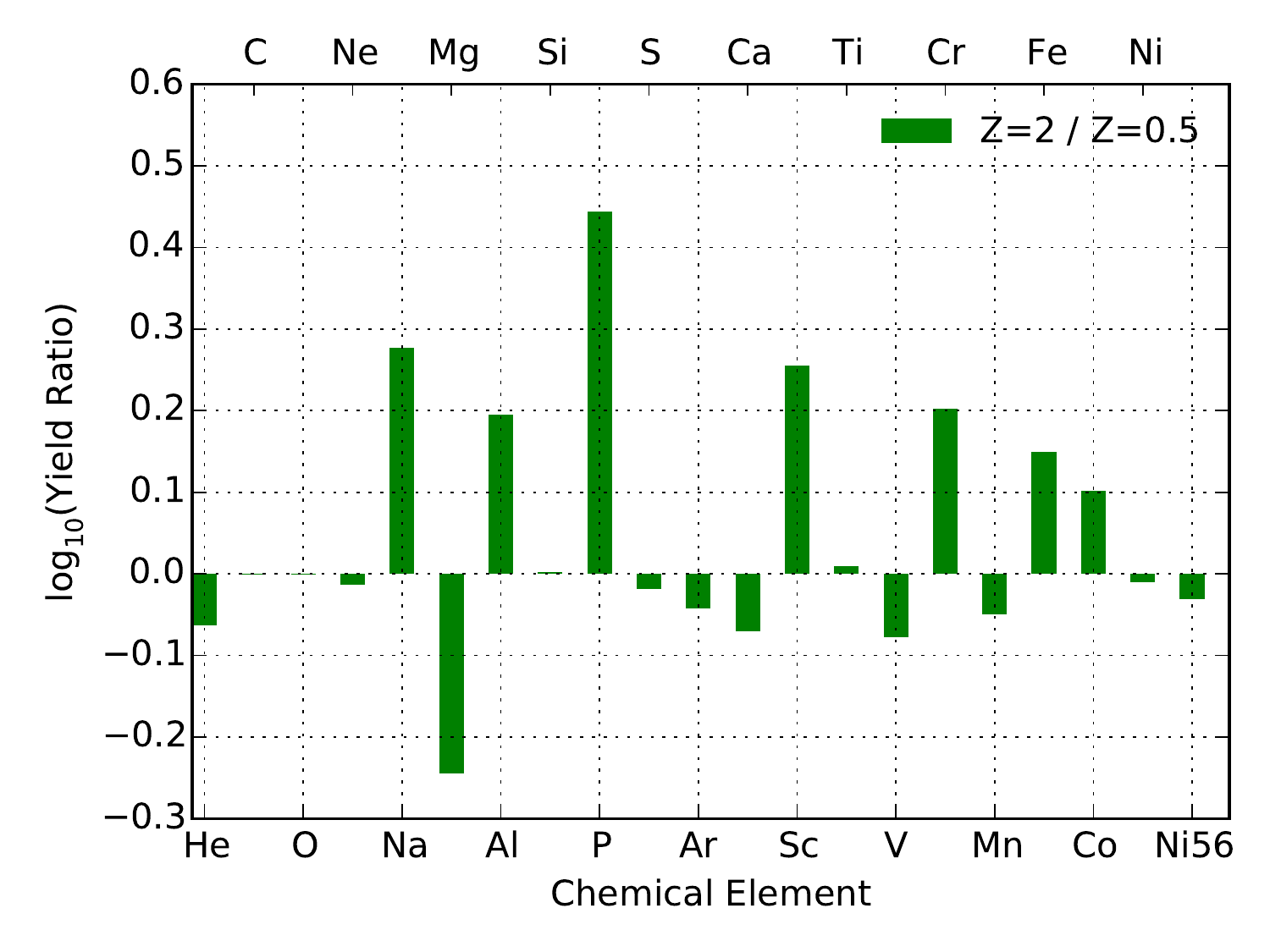}}
\centerline{\includegraphics[width=.5\textwidth]{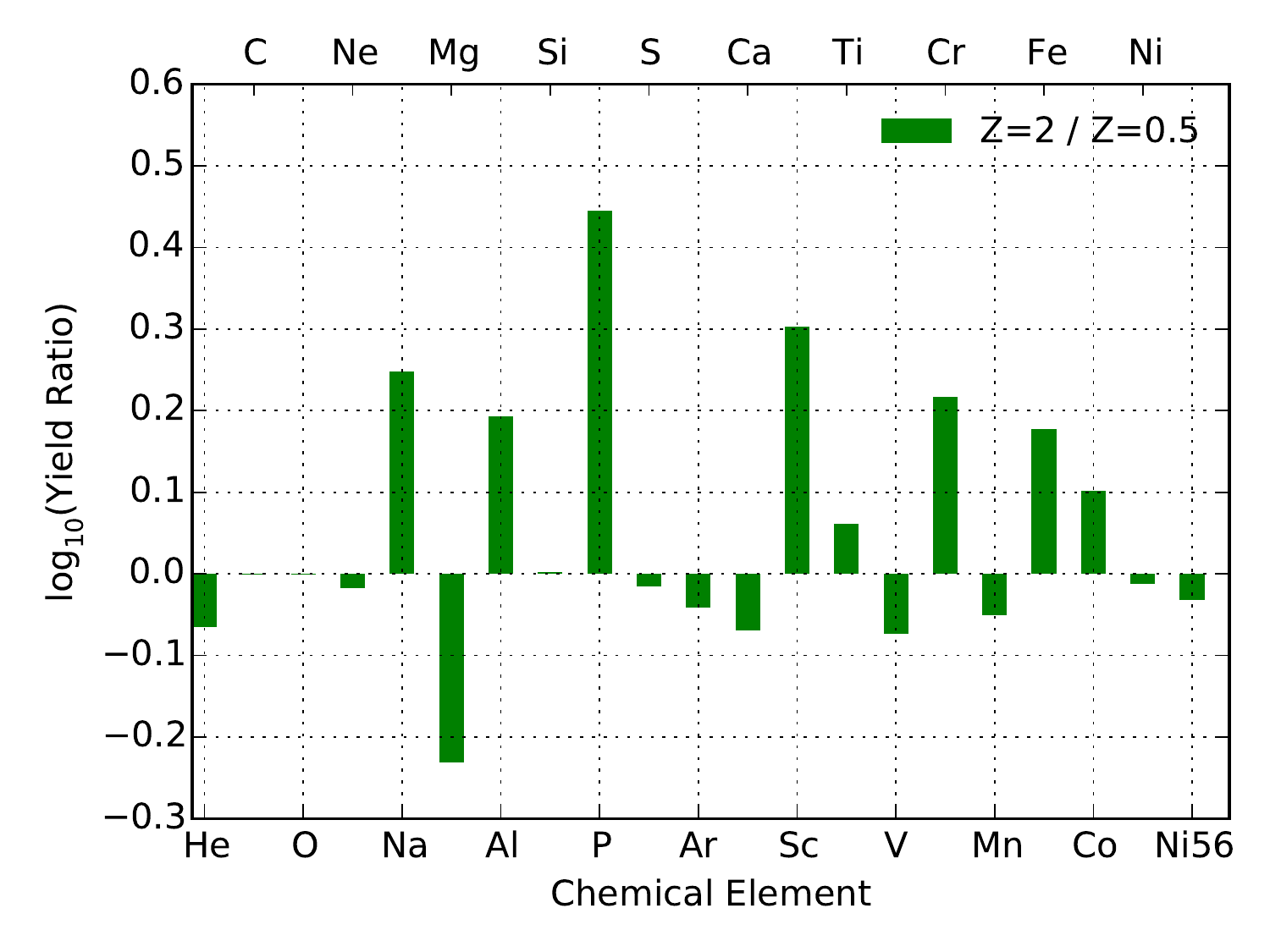}}
\centerline{\includegraphics[width=.5\textwidth]{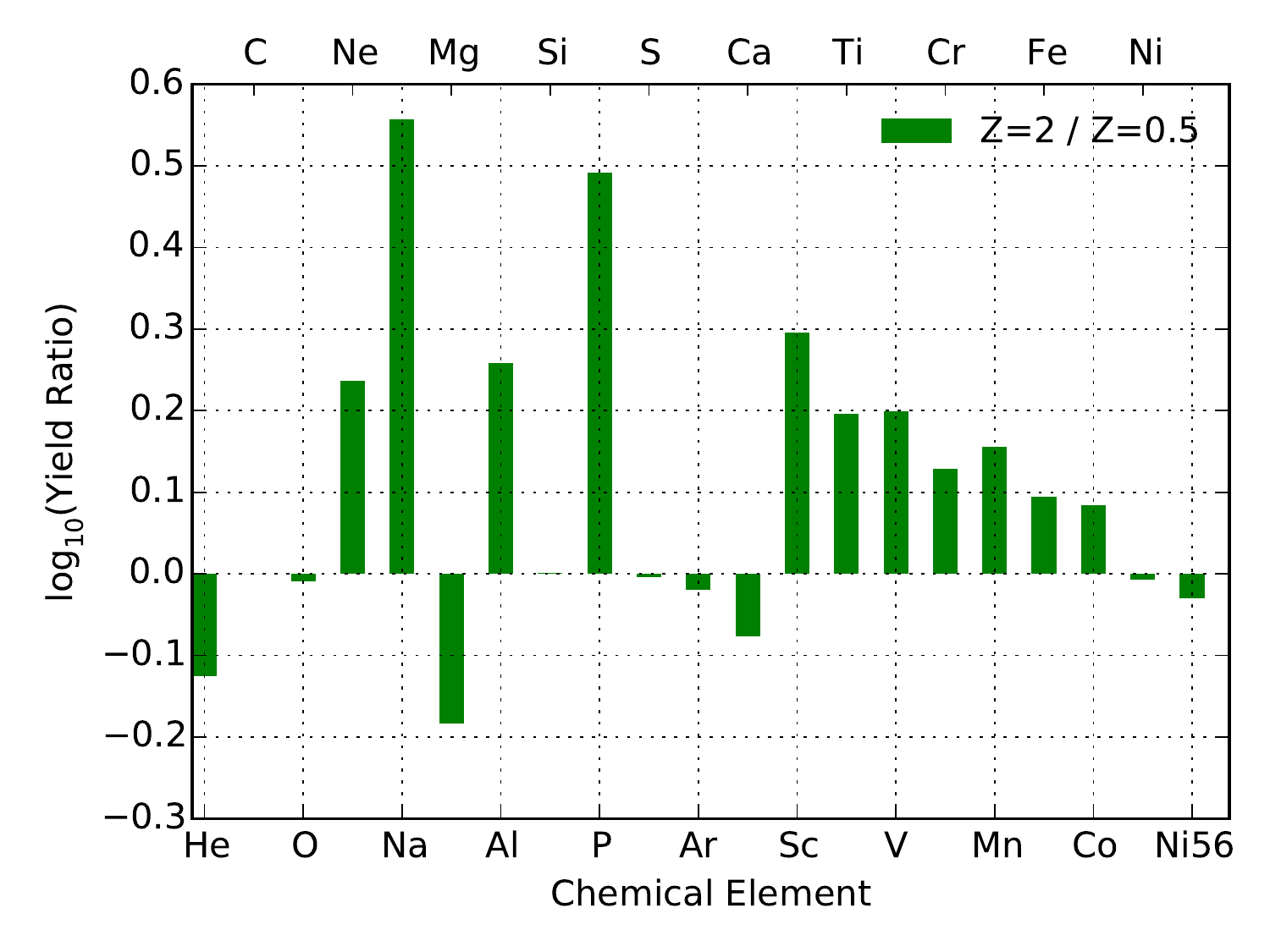}}
\caption{Ratios of the nucleosynthetic yields of high (Z=2 \zsun) versus low (Z=0.5 \zsun) progenitor metallicities for the DDT-high model (top), the DDT-low model (middle), and W7 (bottom) on a logarithmic scale.
 A value close to 0, like Si, means that the same nucleosynthetic yields were obtained for different progenitor metallicities.
 The effect of progenitor metallicity on the total nucleosynthetic yields of the explosion are fairly similar between the two DDT models. The affects on W7 are fairly consistent in sign with the DDT models except for a number of elements above scandium.
}\label{fig:abundance_ratios}
\end{figure}

The relative change in abundances with progenitor metallicity is plotted in Figure~\ref{fig:abundance_ratios}.
The effect of progenitor metallicity on the nucleosynthetic yields of the explosion are fairly similar among the three models for most elements, especially for the DDT-high and DDT-low models.
The most marked difference between W7 and the 2D DDT models is that the Ne and Na yields increase more with metallicity in the W7 model.

\begin{figure}
\centerline{\includegraphics[width=.5\textwidth]{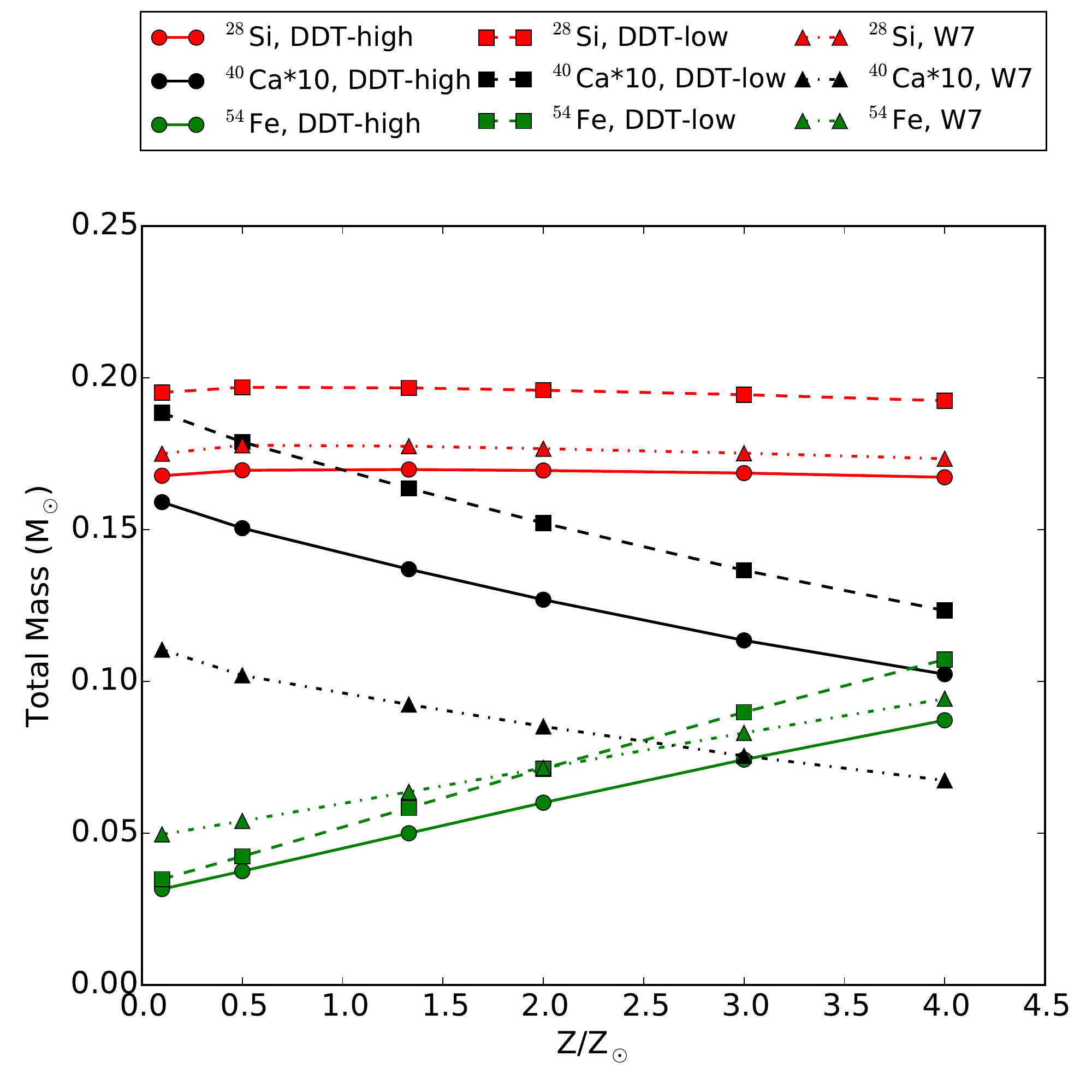}}
\caption{Total yields in M$_\odot$ of $^{28}$Si (red), $^{54}$Fe (green), $^{40}$Ca (black) as a function of metallicity for the two realizations, DDT-high (circles, solid line) and DDT-low(squares, dashed lines), as well as W7 (triangles,dash dot lines)}
\label{fig:total_yields}
\end{figure}

\begin{figure}
\centerline{\includegraphics[width=.49\textwidth]{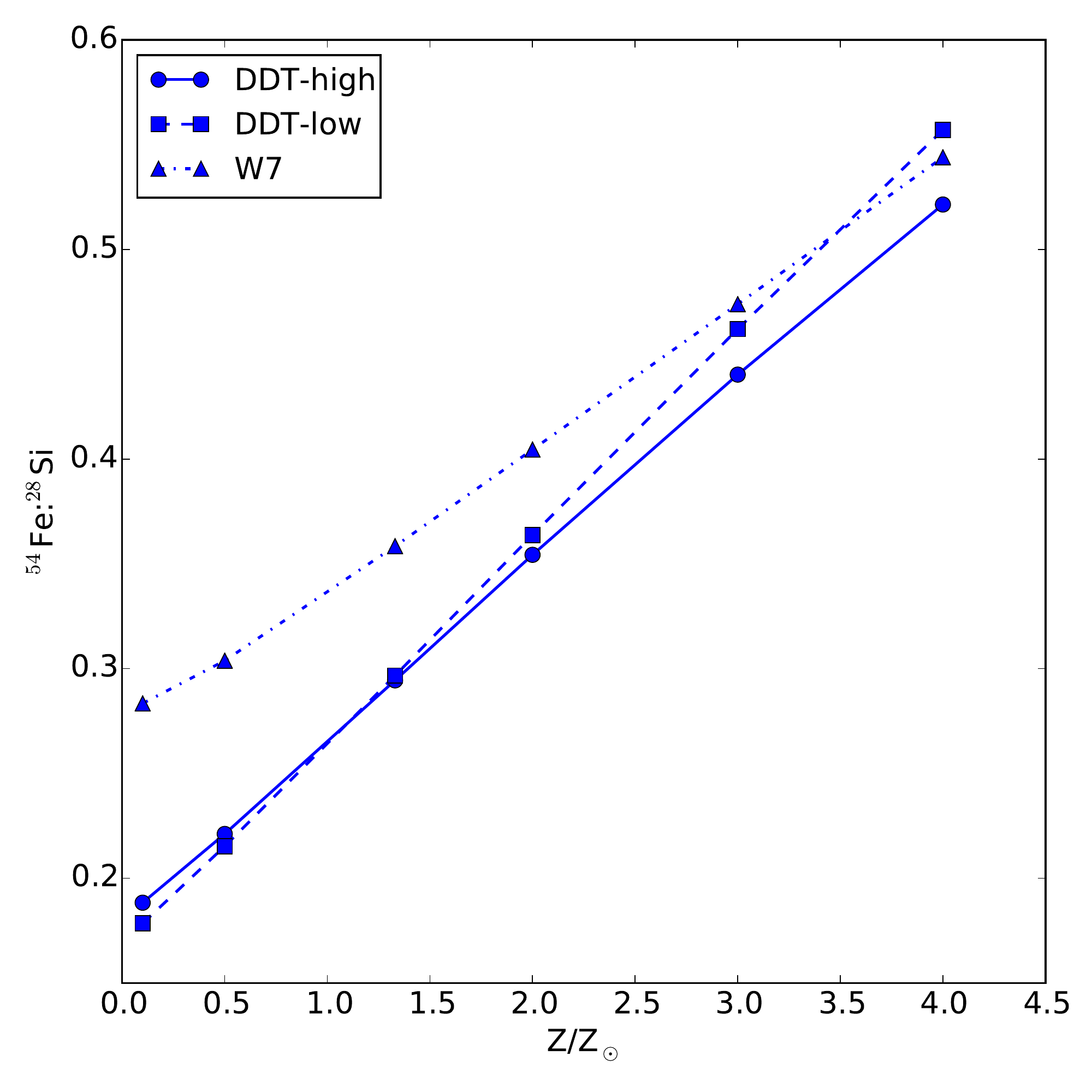}}
\centerline{\includegraphics[width=.5\textwidth]{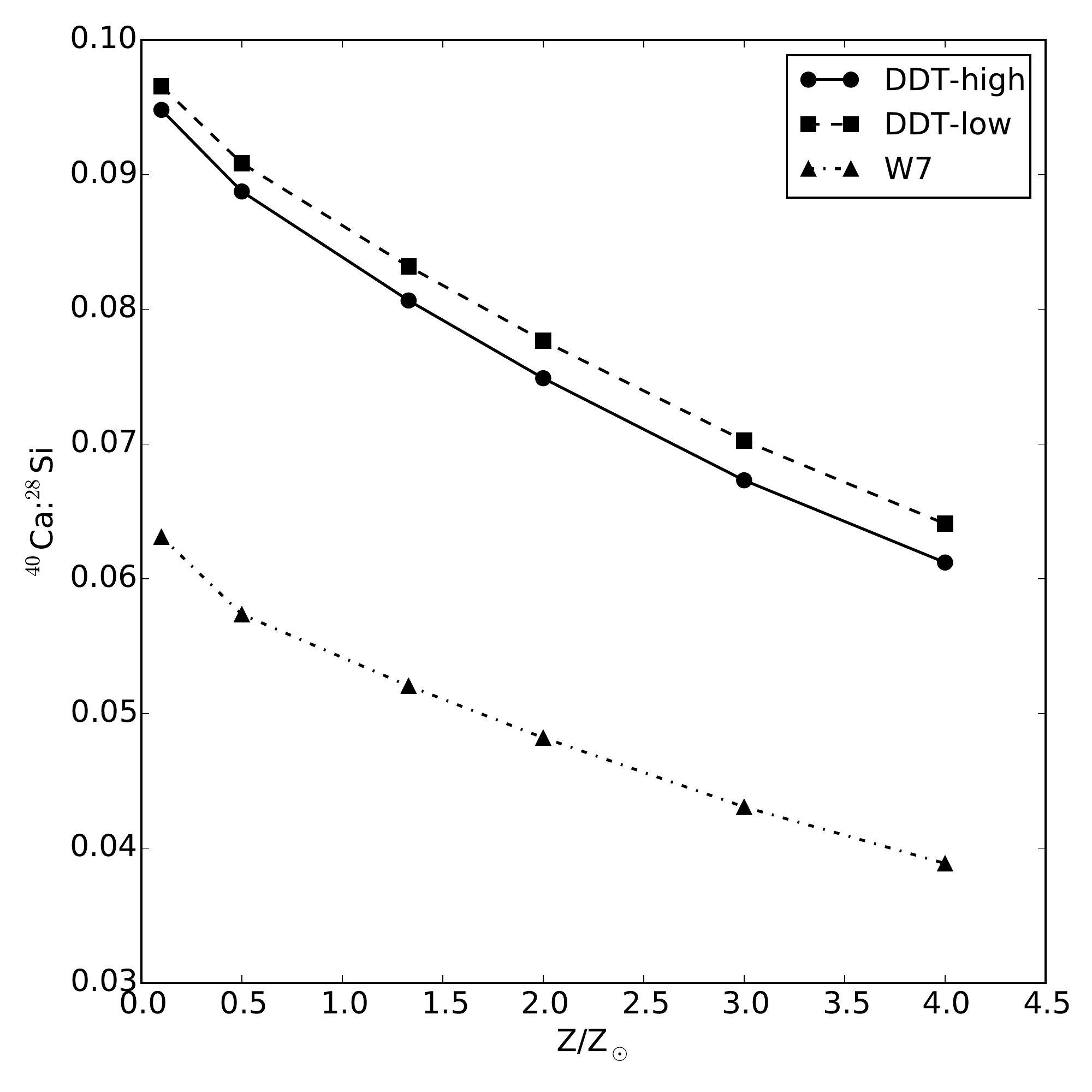}}
\caption{Ratio of $^{54}$Fe to $^{28}$Si yields (top) and $^{40}$Ca to $^{28}$Si yields (bottom) from the 2D DDT and the W7 simulations.
The ratios of these abundances show similar smooth trends with progenitor metallicity for these three models.}
\label{fig:si_ratios} 
\end{figure}

\begin{figure*}
\centerline{\includegraphics[width=.5\textwidth]{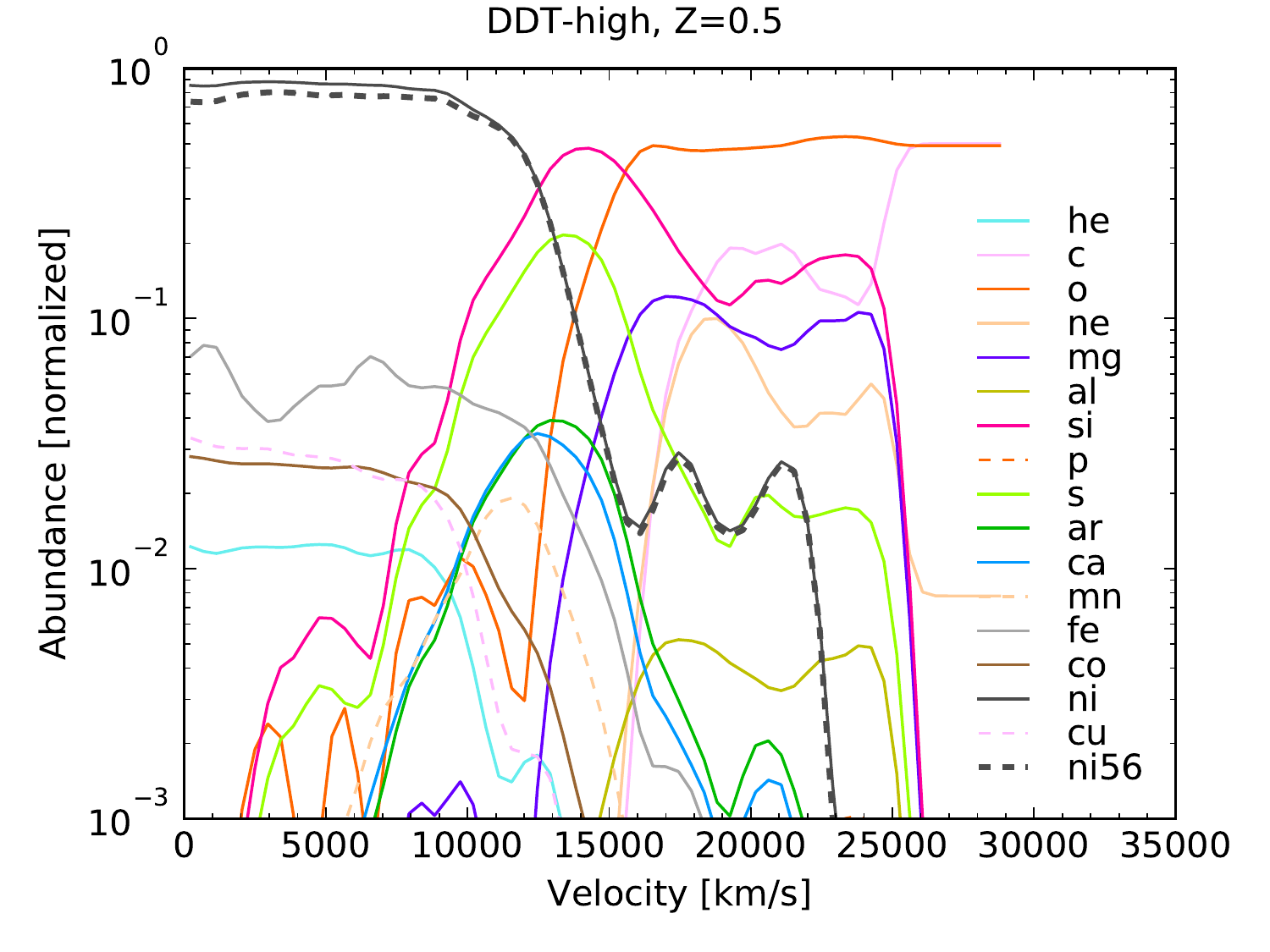}
 \includegraphics[width=.5\textwidth]{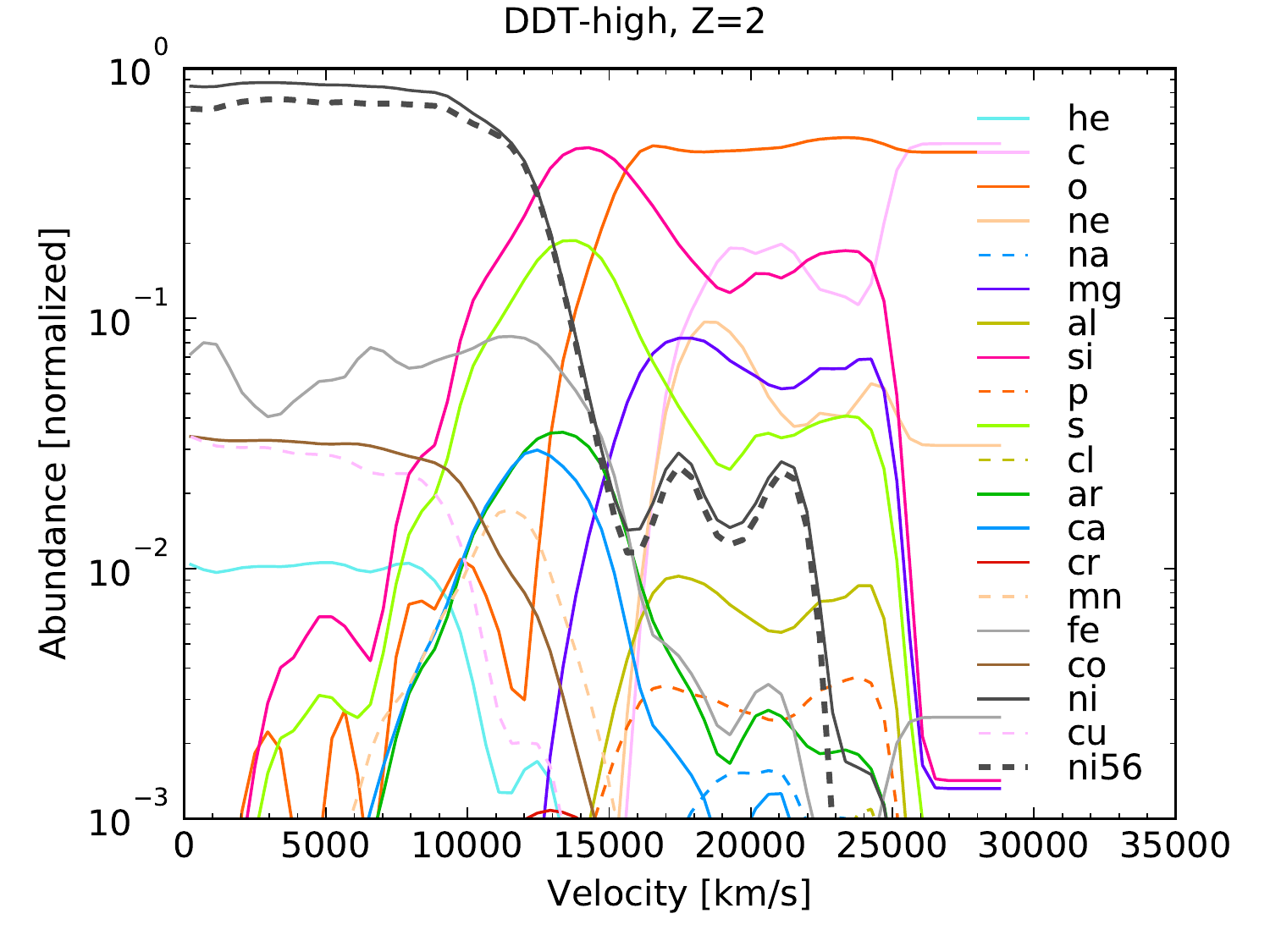}}
\centerline{\includegraphics[width=.5\textwidth]{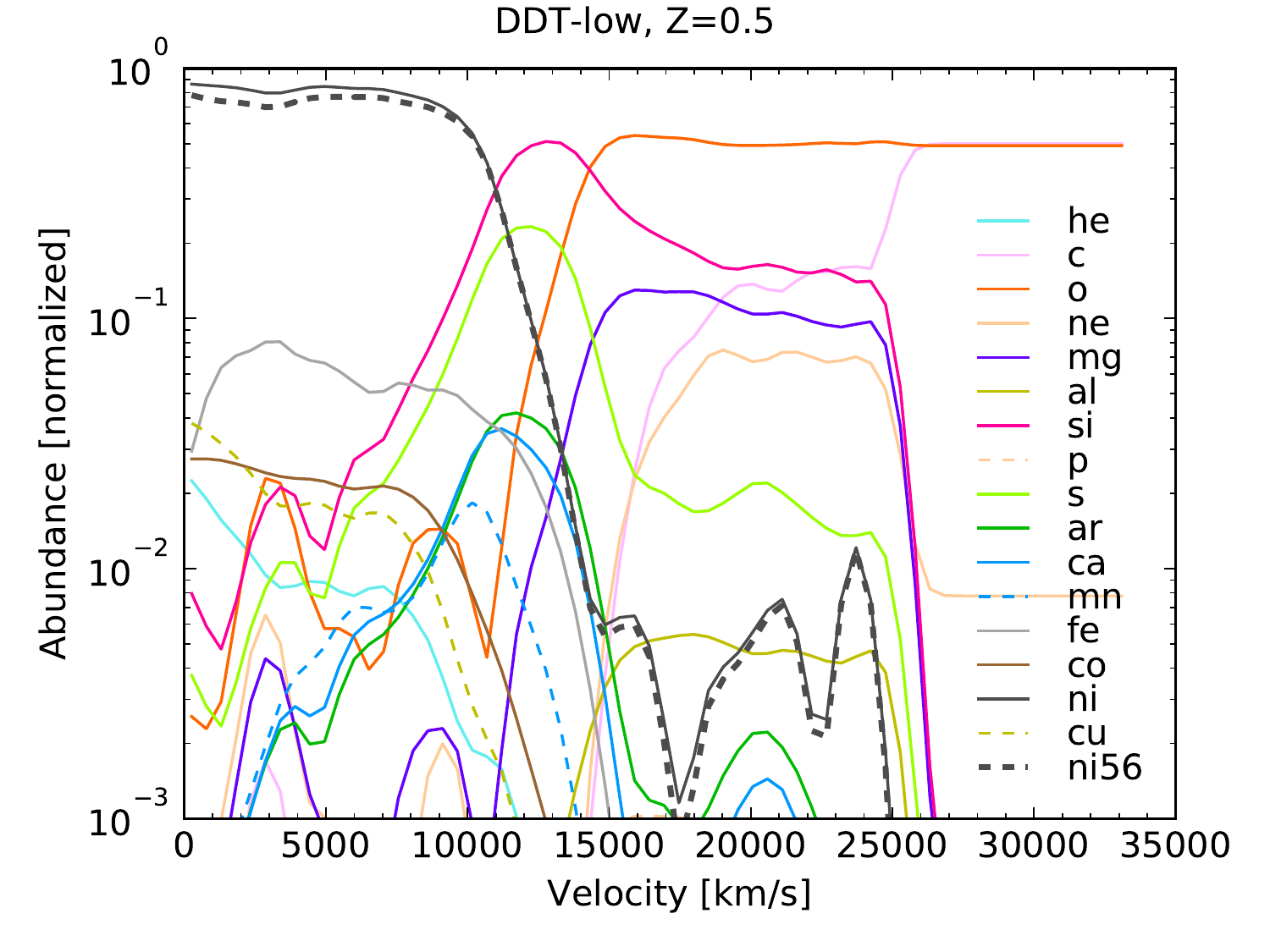} 
 \includegraphics[width=.5\textwidth]{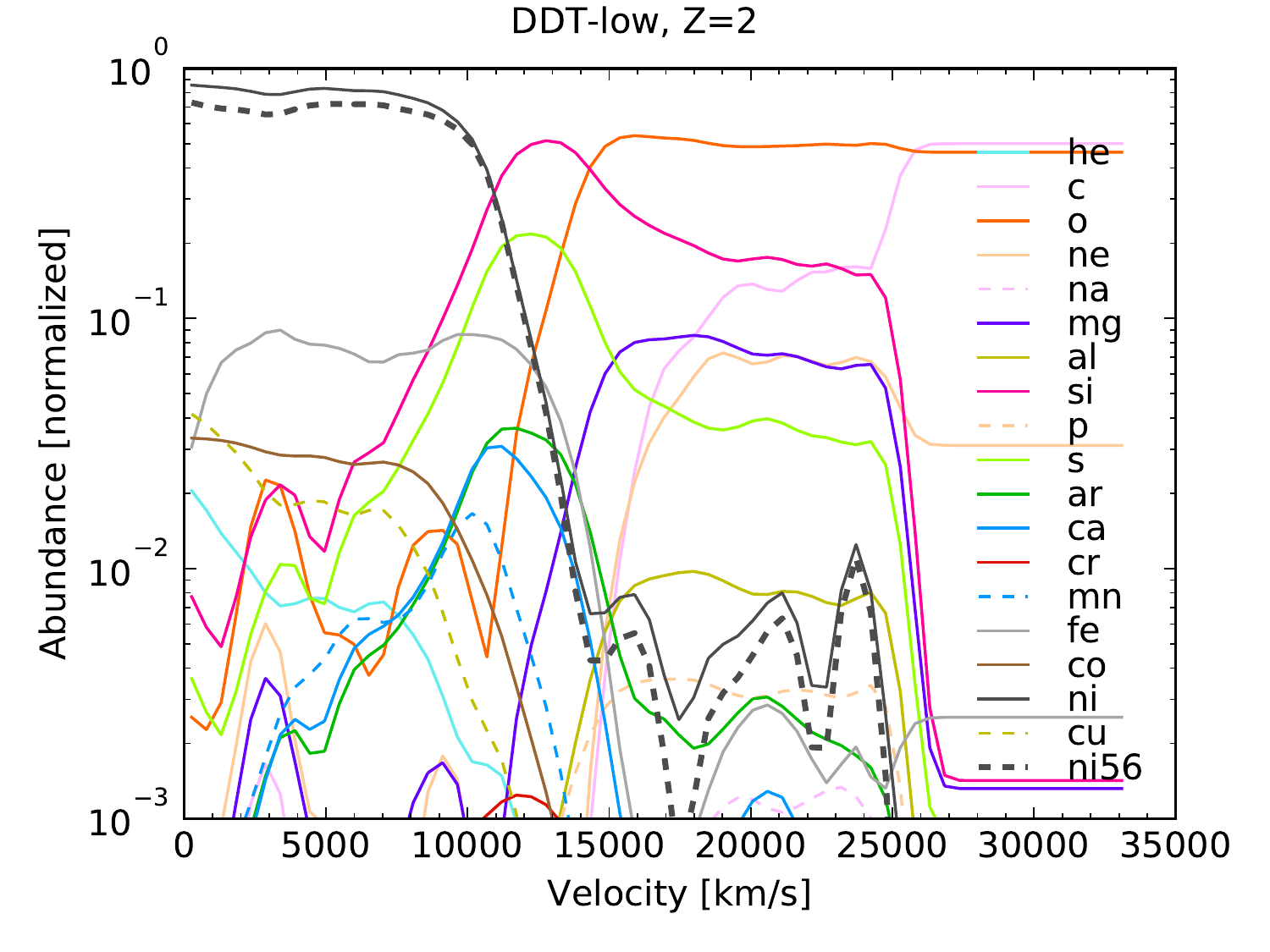}}
\centerline{\includegraphics[width=.5\textwidth]{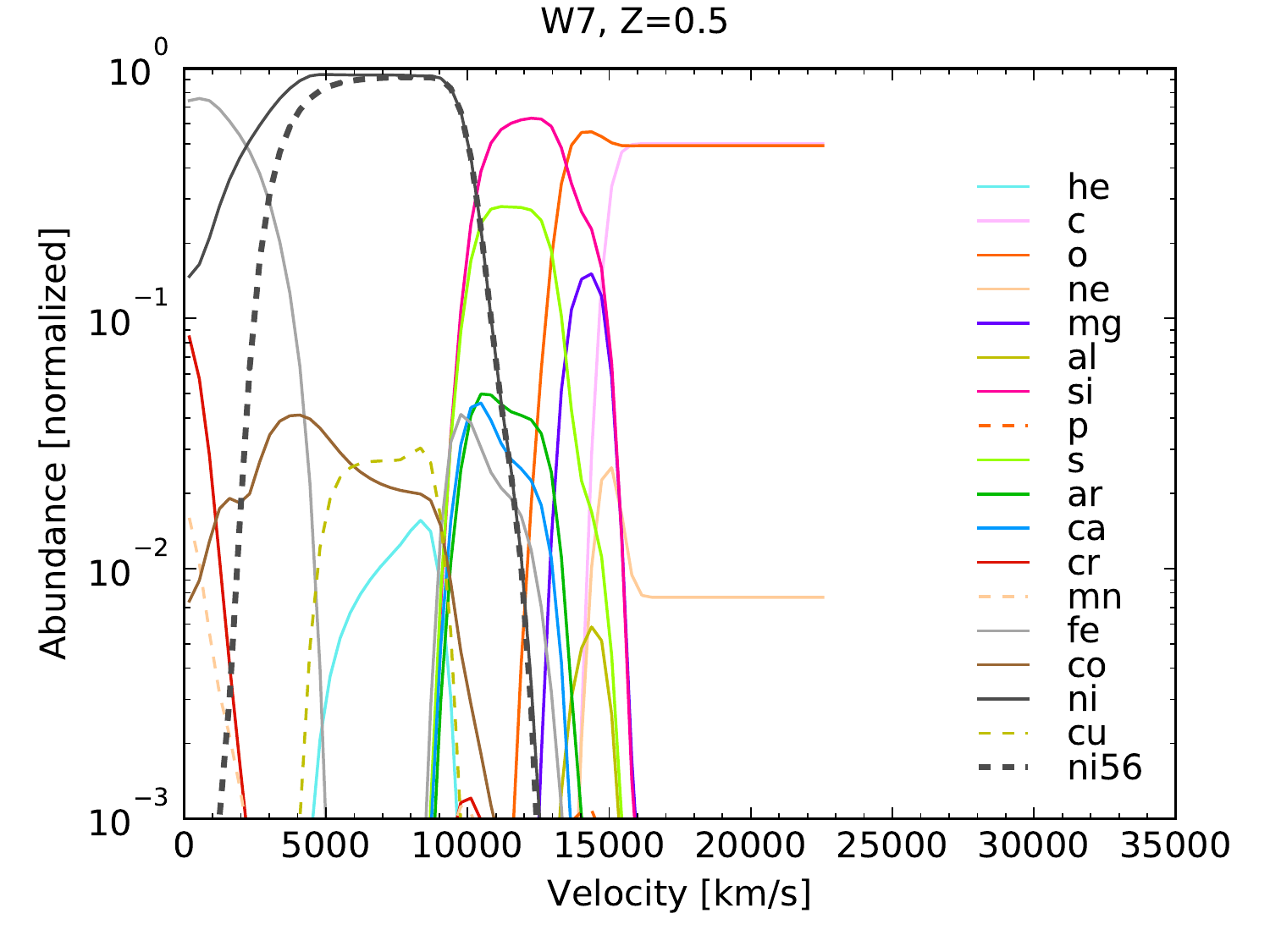}
 \includegraphics[width=.5\textwidth]{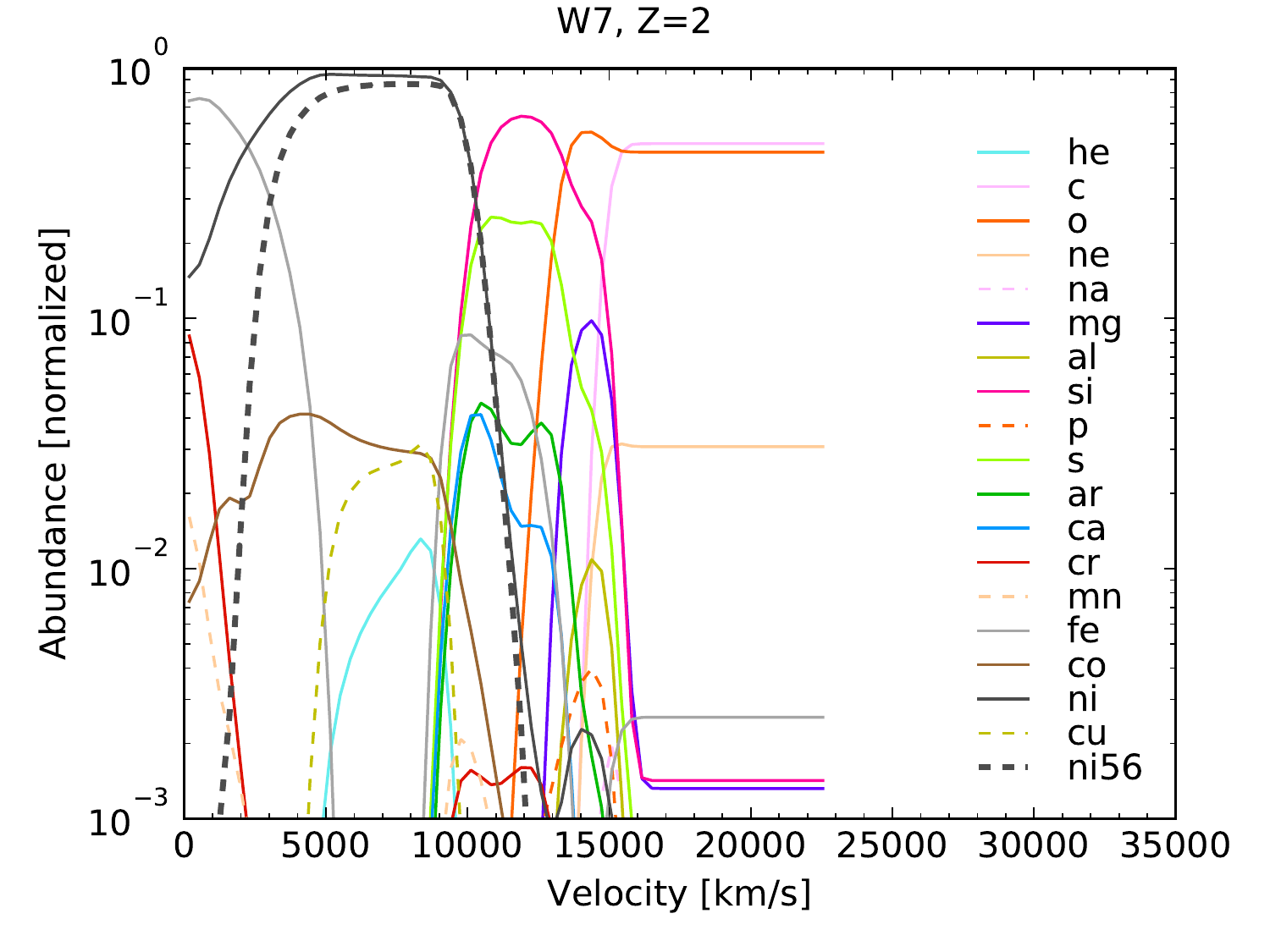}}
\caption{Chemical abundance profiles in 1D spherical velocity space of the DDT-high model (top row), the DDT-low model (middle row), and the W7 model (bottom row) for two different initial metallicities, Z=0.5 and Z=2.
}\label{fig:abundance_profiles}
\end{figure*}

Figure~\ref{fig:total_yields} shows the integrated yields of $^{28}$Si, $^{40}$Ca, and $^{54}$Fe for the six progenitor metallicities.
We find that with increasing metallicity, the overall yields of $^{28}$Si stay relatively constant.
Conversely, the total yields of $^{40}$Ca decrease and $^{54}$Fe increase with increasing metallicity similar to the trends seen in \cite{De14} with decreasing Y$_e$.
The relatively flat trend in the $^{28}$Si yields makes this species a useful reference for comparing the trends in the yields of other species between the three models.
The ratios of \caft and \feff relative to \site are shown in Figure~\ref{fig:si_ratios}.
As expected from Figure~\ref{fig:abundance_ratios}, the trends are similar among the models, though the absolute ratio of $^{40}$Ca to $^{28}$Si is different in the W7 model due to Ca being produced in higher amounts at a wider extent in the DDT models (Figure~\ref{fig:abundance_profiles}).
The trends are also smooth over the metallicity range that we are studying.

Figure~\ref{fig:abundance_profiles} shows the chemical abundance profiles in velocity space from the DDT-high, DDT-low, and W7 models for two metallicities: Z=0.5 and 2.0 \zsun.
Differences in explosion model produce differences in the profiles for a single metallicity.
While all three models feature a \nifs core that extends to approximately $1.0\times10^9$ cm/s, W7 is the only model that has a ``nickel hole'', which is a known characteristic of 1-D models.
In the higher velocity portion of the ejecta, from $\approx$
$1.6\times10^{9}$ cm/s to $2.5\times10^{9}$ cm/s, the W7 model's ejecta consist of unburnt C, O, and Ne, much simpler than that produced in the DDT-high
and DDT-low models. In this region of the material, the DDT-high and DDT-low models'
profiles consist of predominantly C and O, with $\lesssim 10\%$ of Si, Mg, Ne,
S, Ni, Ar, and Ca. At the highest velocities, $\gtrsim 2.7\times
10^9$~cm~s$^{-1}$, the DDT-high and DDT-low ejecta take on a similar configuration to
that of the outer layers of W7, unburnt C, O, and Ne. At intermediate velocities, between $1.0\times10^{9}$ cm/s and $1.6\times10^{9}$, Si peaks in all three models. This is also the location of other IME peaks such as S, Ar, and Ca. 

Examining the same explosion models at different metallicites presents
additional features. At Z = 2 \zsun, \nifs makes up less of the total
Ni yield in the three explosion models, though there is little change in the
overall mass fraction of Ni. Overall, the Fe mass fraction increases with the
increased metallicity. However, not only is the mass fraction of Fe higher,
the Fe rich layers extend further outward into the higher velocity ejecta. In
all three modes, the Fe rich layer reaches into the material that is mostly
intermediate mass elements. The increase in metallicity also causes the
Ca peak to be both lower and narrower. The Si peak is largely unaffected by
the increase in metallicity.  This is also reflected in how little the overall
Si abundance varies in Figure~\ref{fig:total_yields}, and the small change in the Si\,\textsc{ii}
spectral feature at 6150\,\AA\ with metallcity as shown in
section~\ref{sec:spectra}.

\begin{figure}
\centerline{\includegraphics[width=.5\textwidth]{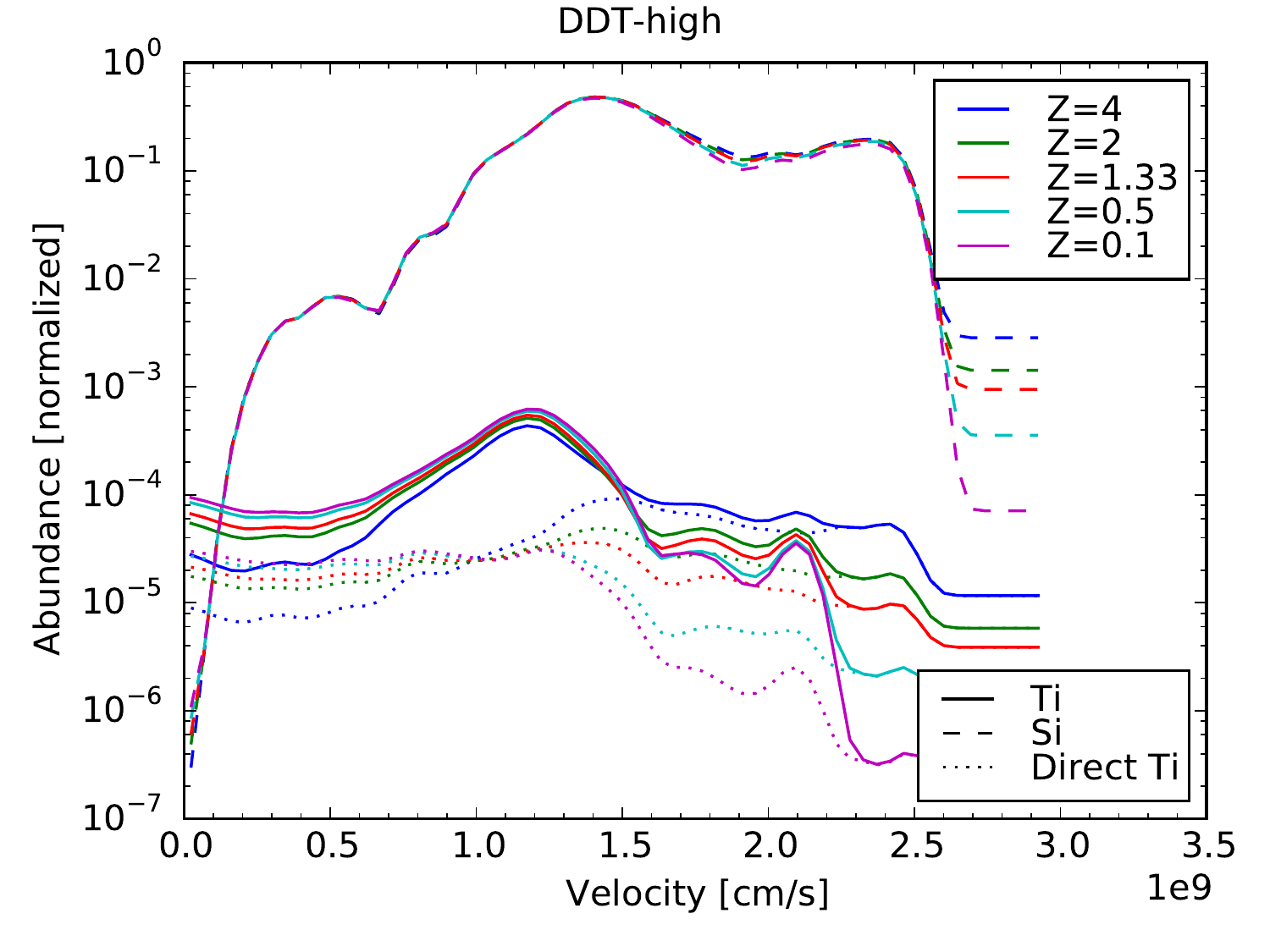}}
\centerline{\includegraphics[width=.5\textwidth]{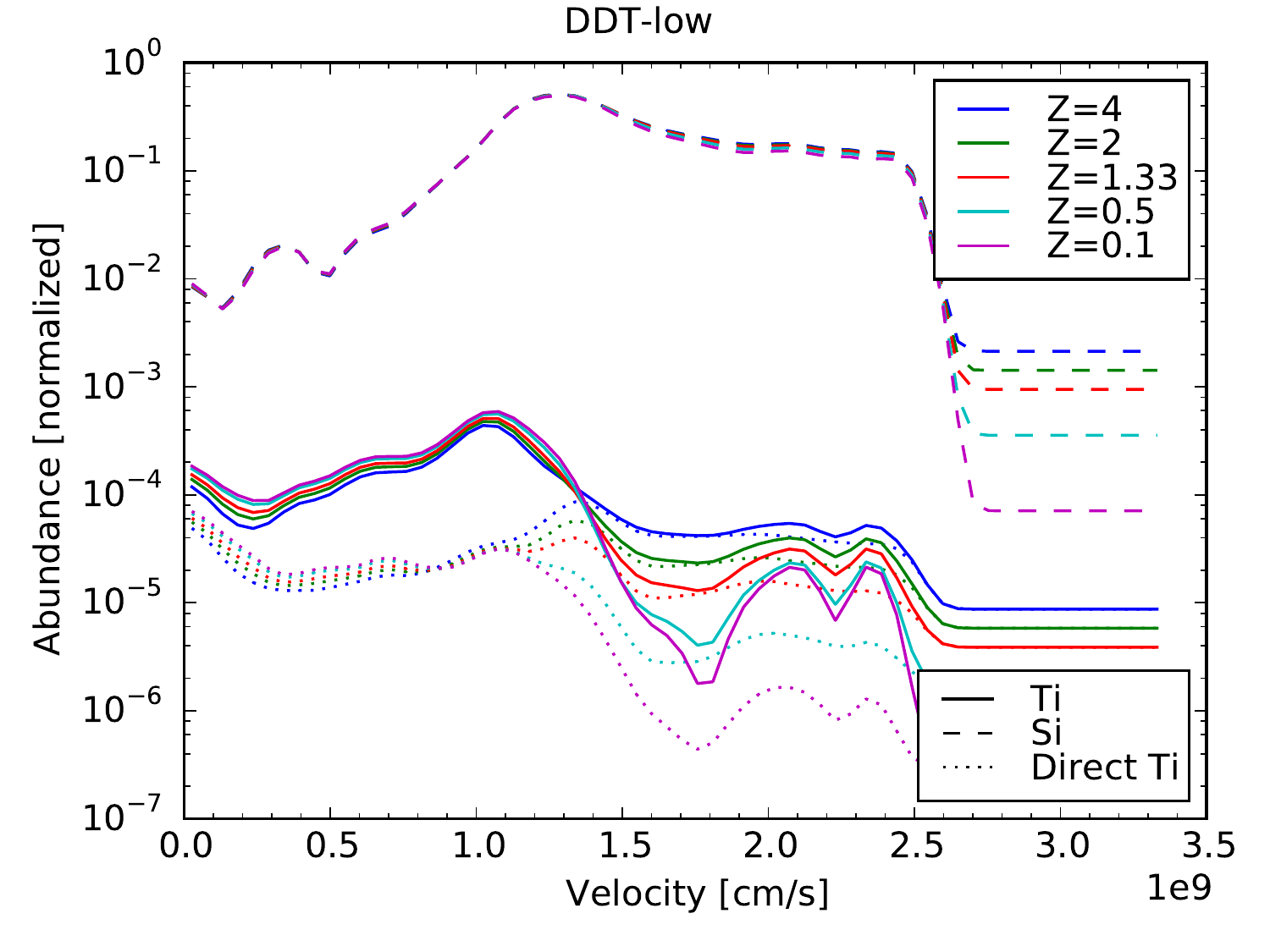}}
\centerline{\includegraphics[width=.5\textwidth]{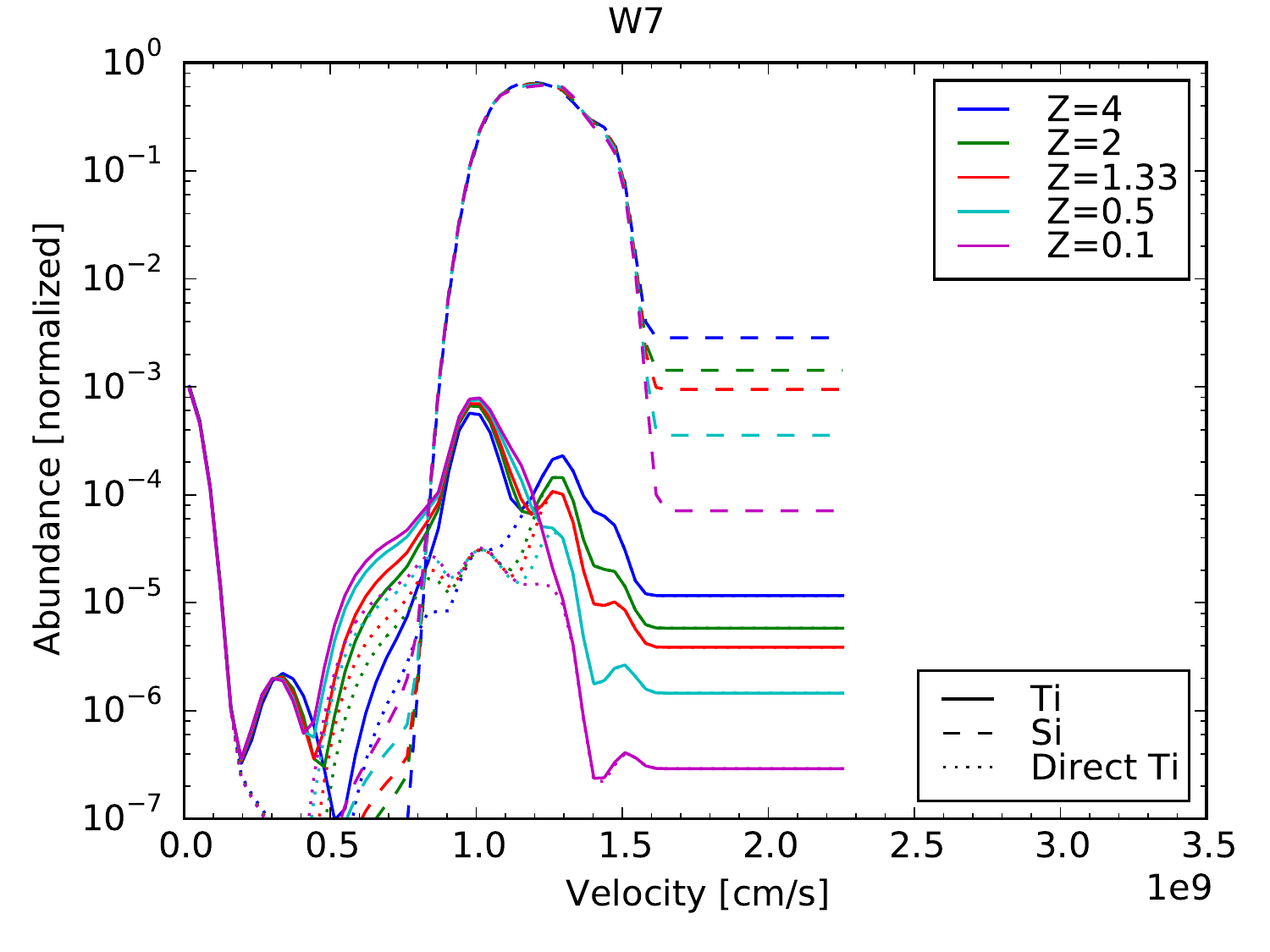}}
\caption{Radial abundance profiles of Si and Ti at day 30 post explosion in the DDT-high model (top), DDT-low model (middle), and the W7 model (bottom).
 Different line colors represent different initial metallicities.
 The Ti yield (solid line) in the outer regions of the burning zone is sensitive to the progenitor metallicity, while the Si production (dashed line) is largely unaffected.
 Also shown, for each metallicity, is the portion of the Ti that is not a product of $^{48}$Cr decay (direct Ti, dotted lines).
}\label{fig:ti_profiles}
\end{figure}

Figure~\ref{fig:ti_profiles} shows the abundance profiles of Si and Ti in the DDT and W7 models 30 days post explosion.
It demonstrates how the Ti production is sensitive to the progenitor metallicity while the Si production is largely unaffected.
The variation of Ti with metallicity is strongest in the region of incomplete Si burning and the unburned regions.
A shift in the quasi-equilibrium abundances due to metallicity is expected in the Si-burning region \citep{De14}.
The direct Ti production is mainly in the form of $^{44}$Ti, which has a half-life of 60 years\footnote{\label{note2}http://www.nndc.bnl.gov/chart/}, while the progenitor metallicity contributes mainly $^{48}$Ti.
Any $^{48}$Cr produced during the explosion will decay with a half-life of 21.56 hours$^{\ref{note2}}$ to $^{48}$V, which will decay further with a half-life of 15.97 days$^{\ref{note2}}$ to stable $^{48}$Ti.
We show the profile at day 30 post explosion to facilitate discussion of Ti spectral features at this epoch in Section~\ref{sec:spectra}.
At this time, 71\% of the $^{48}$Cr produced during the explosion has decayed to $^{48}$Ti so that the Ti curves in Figure~\ref{fig:ti_profiles} in some regions are higher than the ``direct Ti'' curves.

\subsection{Light Curves}

\begin{figure}
\centerline{\includegraphics[width=.5\textwidth]{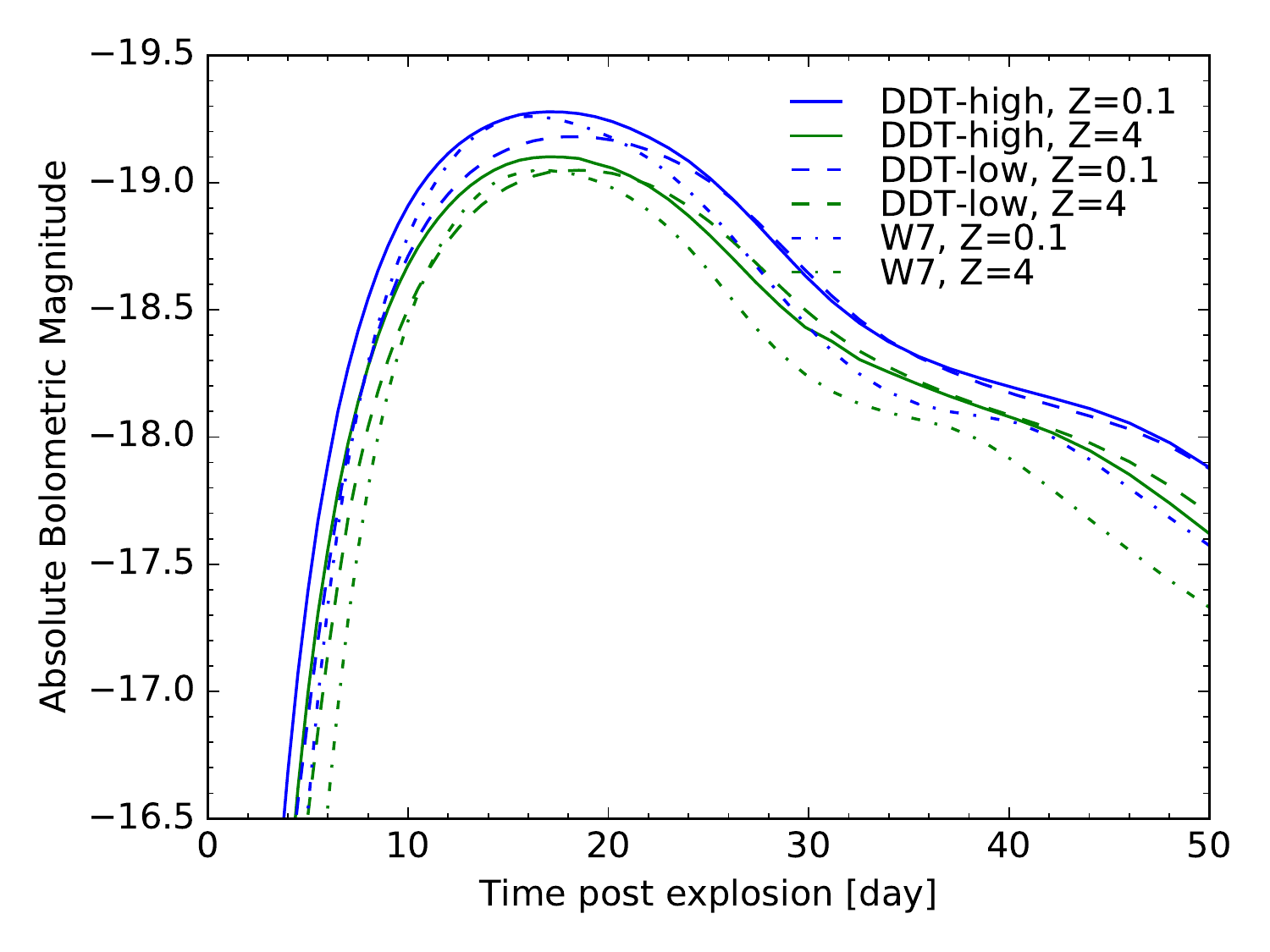}}
\caption{Bolometric light curves of the DDT-high (solid lines), DDT-low (dashed lines), and W7 (dotted lines) explosion simulations at two extreme progenitor metallicities of 0.1 (blue lines) and 4 times solar (green lines).
 The light curves of explosion simulations with high metallicity progenitor are brighter at peak than those with low metallicity progenitors because the former have a higher \nifs\ yield.
}\label{fig:lc0}
\end{figure}

The bolometric light curves for the three explosion models with two extreme metallicities are shown in Figure~\ref{fig:lc0}.
Higher progenitor metallicities systematically make the bolometric light curve rise more slowly and peak lower.
The slower rise is caused by the higher abundance of metals in the outer layers of the ejecta, which raise the opacity, and make radiation diffuse out more slowly from the hot \nifs-rich core.
The lower peak brightness is caused by the lower total \nifs mass in the models with higher progenitor metallicities, as shown in Table~\ref{tab:yields_grouped}.

\begin{figure}
\centerline{\includegraphics[width=.5\textwidth]{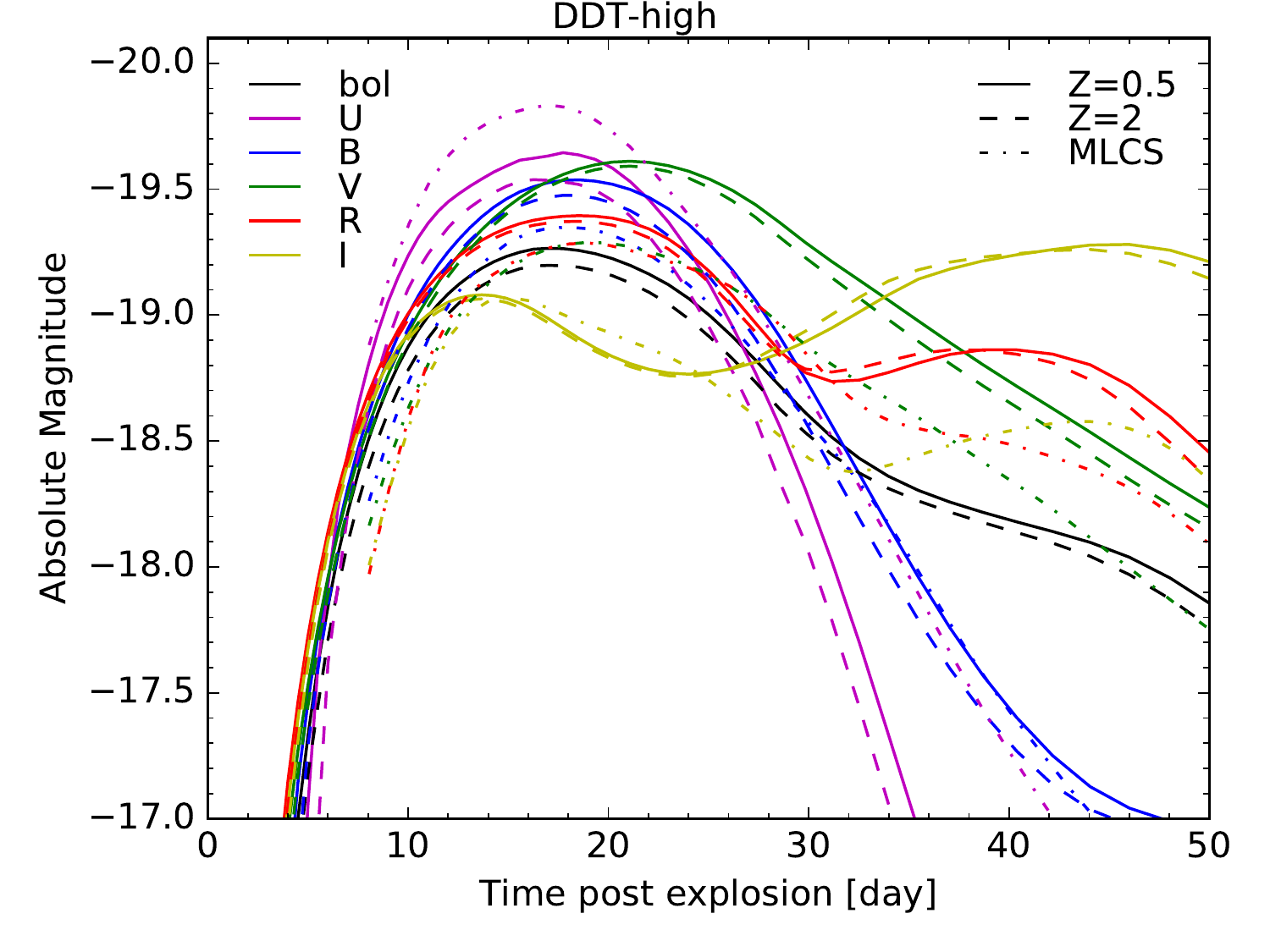}}
\centerline{\includegraphics[width=.5\textwidth]{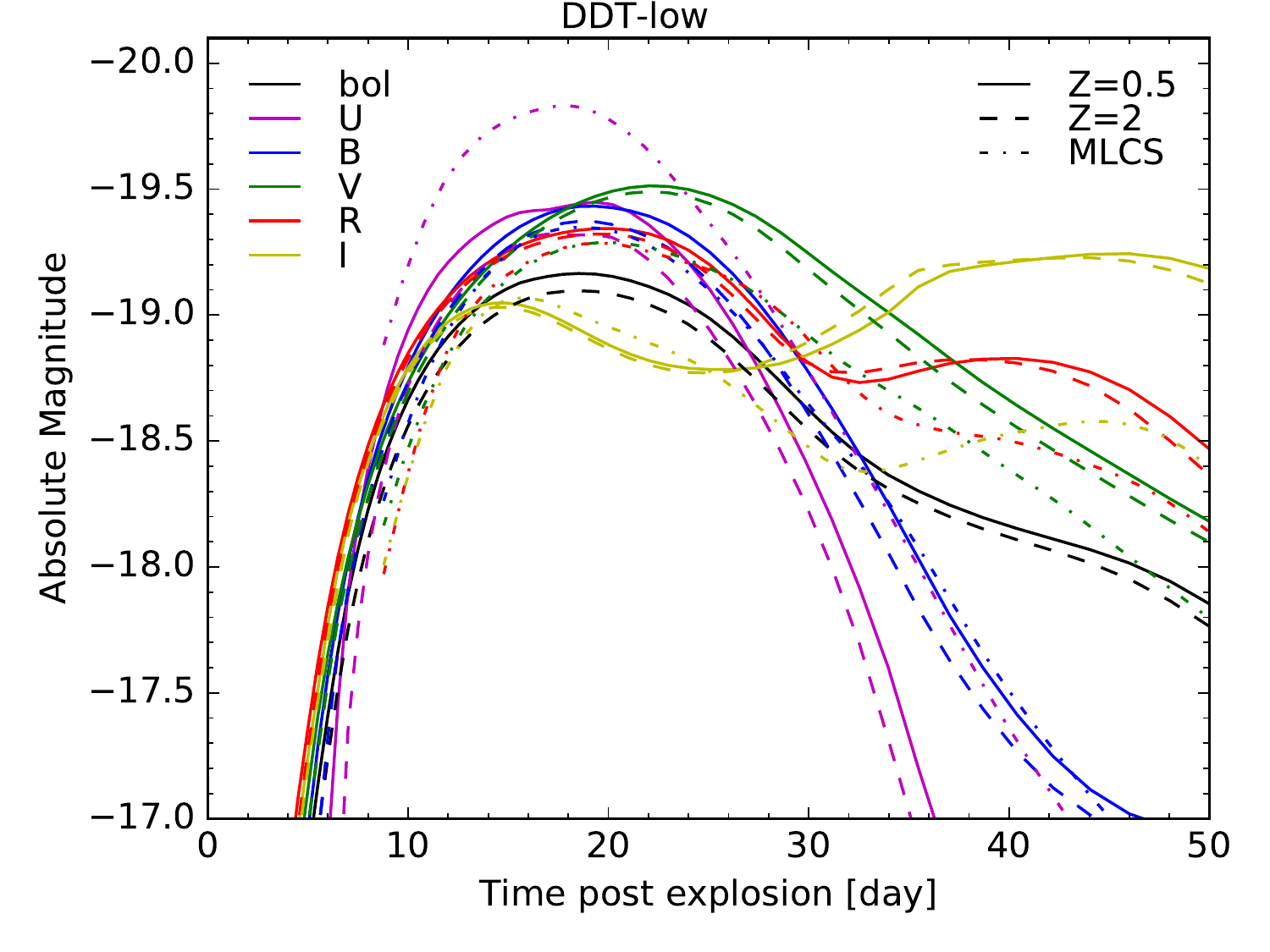}}
\centerline{\includegraphics[width=.5\textwidth]{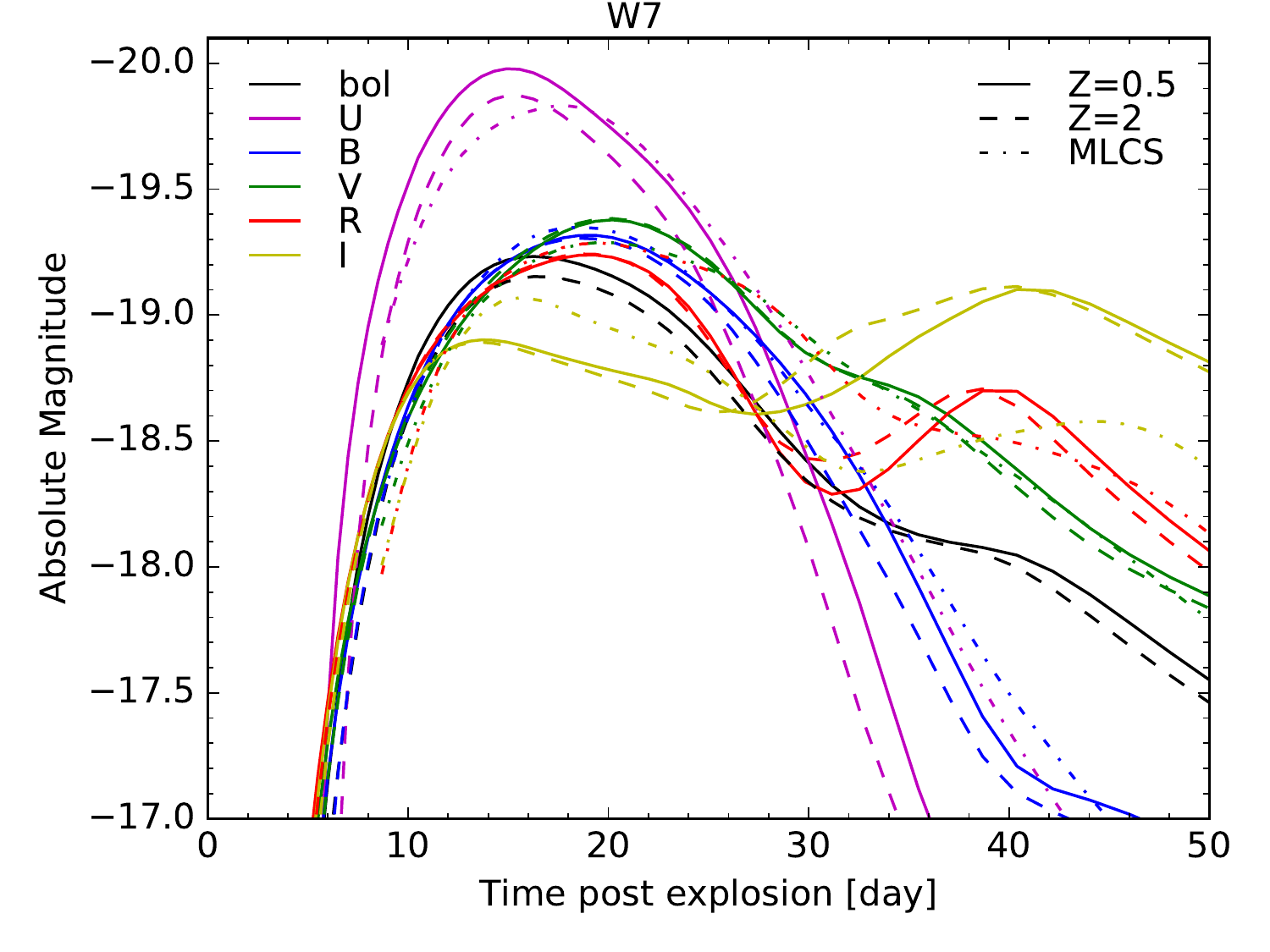}}
\caption{Bolometric and UBVRI-band light curves of the DDT-high (top), DDT-low (middle) and the W7 model (bottom), comparing progenitor metallicities of Z=0.5 (solid lines) and Z=2 (dashed lines).
 As a reference, MLCS light curves are plotted for a value of MLCS $\Delta=0$ (dash-dotted lines).
 The models are similar in brightness and color evolution, especially the DDT-high and DDT-low simulations.
 The biggest difference is the peak brightness in the U-band between the 2D DDT simulations and W7.
}\label{fig:lc}
\end{figure}

\begin{figure}
\centerline{\includegraphics[width=.5\textwidth]{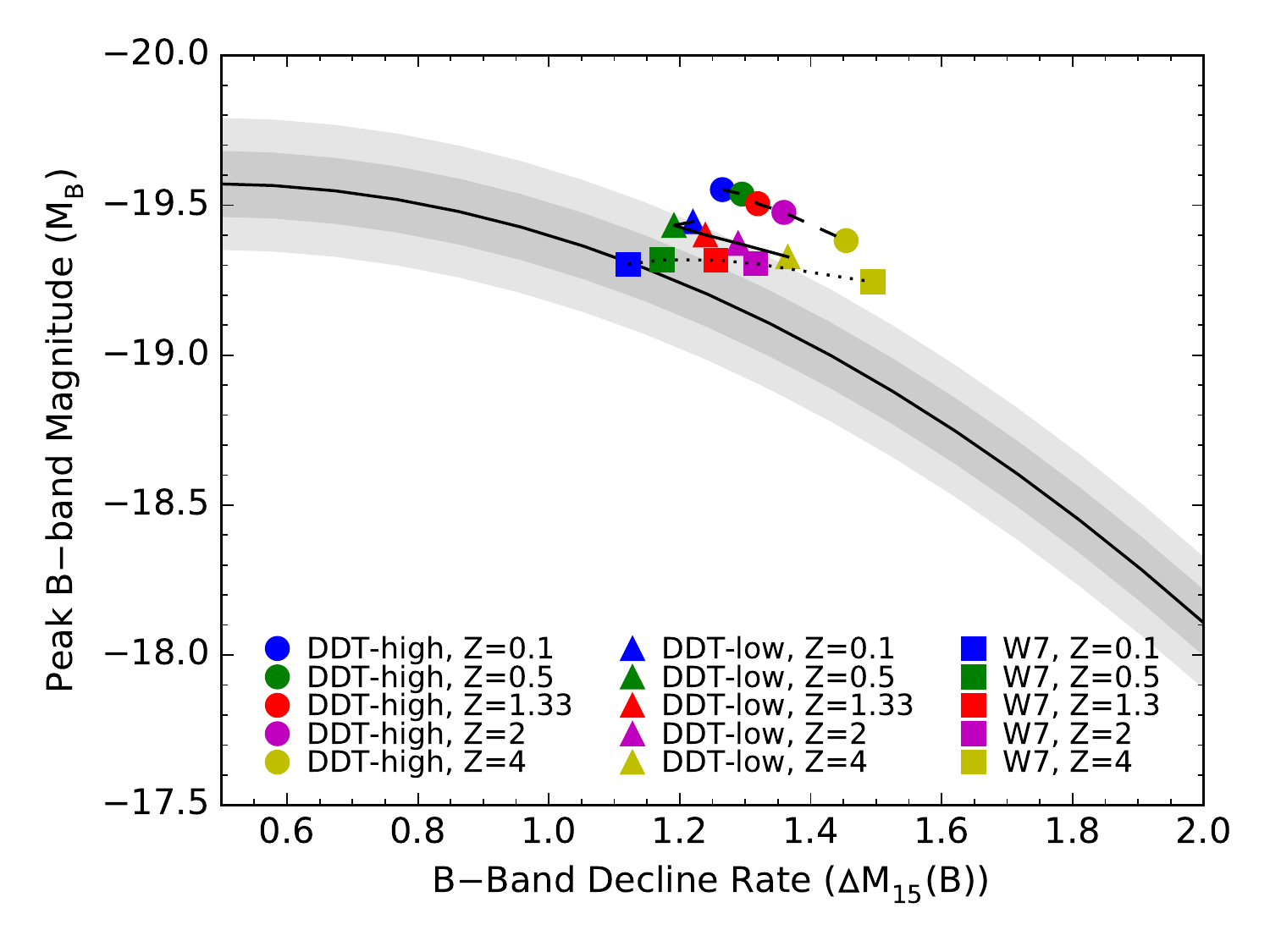}}
\vspace{-10pt}
\centerline{\includegraphics[width=.5\textwidth]{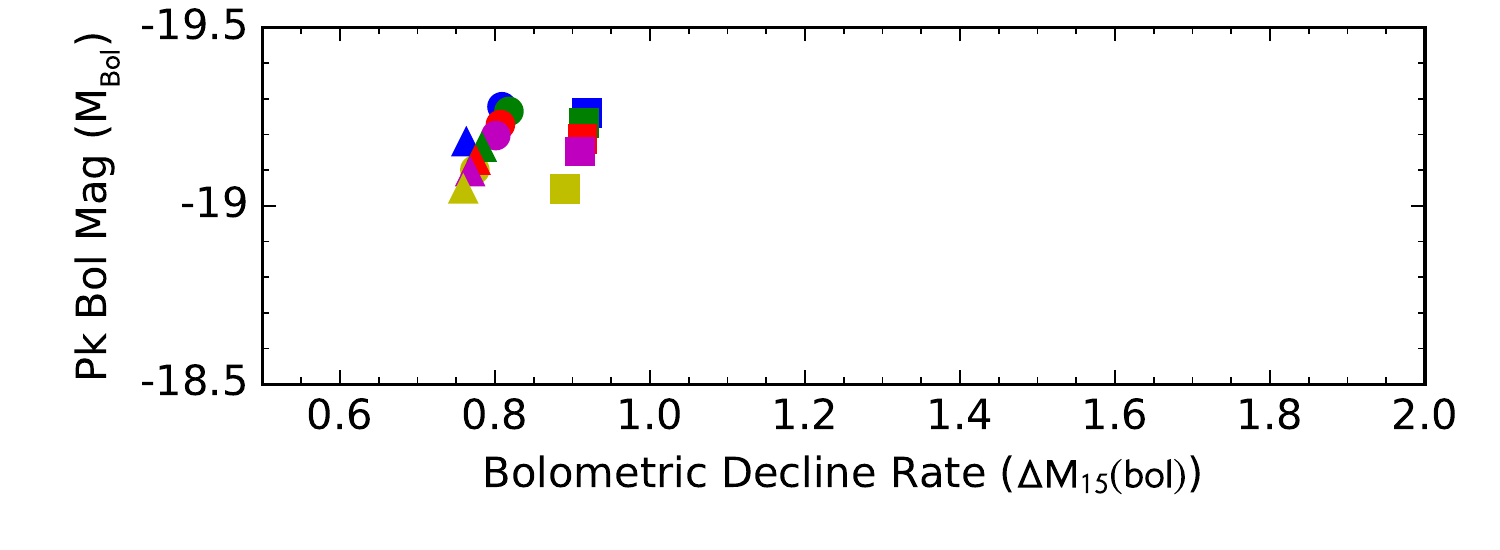}}
\caption{Peak B-band magnitude ($M_{\rm B}$) vs.\ B-band decline rate ($\Delta M_{15}(B)$) (\emph{top panel}) and peak bolometric magnitude vs.\ bolometric decline rate (\emph{bottom panel}), for the DDT-high, DDT-low, and W7 models with initial metallicities of 0.1, 0.5, 1.33, 2, and 4 times solar.
The \citet{Phillips_99} relation is plotted in gray, with the shaded regions indicating the 1 and 2 $\sigma$ observed variation also from that work.
The range in $\Delta M_{15}$ is present in the B-band but not the bolometric light curves showing that this is a color effect, as pointed out by \cite{Kasen2007On-the-Origin-o} 
}\label{fig:phillips}
\end{figure}

In Figure~\ref{fig:lc} shows the UBVRI-band model light curvescompared to the light curves from the data driven model MLCS \cite{Jha07}.
The systematic trend with progenitor metallicity in most light curves is small but significant, especially in the UBV-bands.
This is apparent in Figure~\ref{fig:phillips}, where the light curve shape parameters, peak brightness and decline rate $\Delta$M$_{15}$(B), are plotted, along with the \citet{Phillips_99} relation (see also \citealt{Mandel_11}).
The model light curves have slightly brighter peak magnitudes than the bulk of observed \sneia\ at the same $\Delta m_{15}(B)$.
We find for the two 2D DDT cases used here, which have above-average yields, that the changes in ignition conditions that lead to different $^{56}$Ni yields cause the shape parameters to change perpendicular to the \citet{Phillips_99} relation.
This confirms a similar result found by \citet{Seitenzahl13} using 3D DDT models.
This failure of multidimensional DDT models to reproduce the Phillips relation in the expected way makes it difficult to make comparisons with empirical metallicity-luminosity relations.
In contrast, the progenitor metallicity introduces a change in the B-band light curve shapes that roughly follows the observed width-luminosity relation.

The bottom panel of Figure~\ref{fig:phillips} shows the bolometric peak magnitude and decline rate.
While higher metallicity leads to lower bolometric peak magnitudes, just as it leads to lower $B$-band peak, the effect on $\Delta M_{15}(\rm bol)$ is very small.
This indicates that the range of $\Delta M_{15}(B)$ (top panel of Figure~\ref{fig:phillips}) spanned by metallicity is due entirely to a color effect, in which the B-band light decreases more quickly at higher metallicity while the decline rate of the total energy output is mostly unchanged.
$\Delta M_{15}(\rm bol)$ is even slightly slower at higher metallicity, the opposite of $\Delta M_{15}(B)$.
In reality, any color effects caused by progenitor metallicity would be convolved with environmental factors such as extinction from dust.
Light curve modeling techniques such as MLCS \citep{Riess_MCLS,Jha07} and BayeSN \citep{Mandel_11} are able to separate the two, allowing them to have the possibility of properly capturing the effects of the progenitor metallicity.

\subsection{Spectra} \label{sec:spectra}

\begin{figure*}
\centerline{\includegraphics[width=\textwidth]{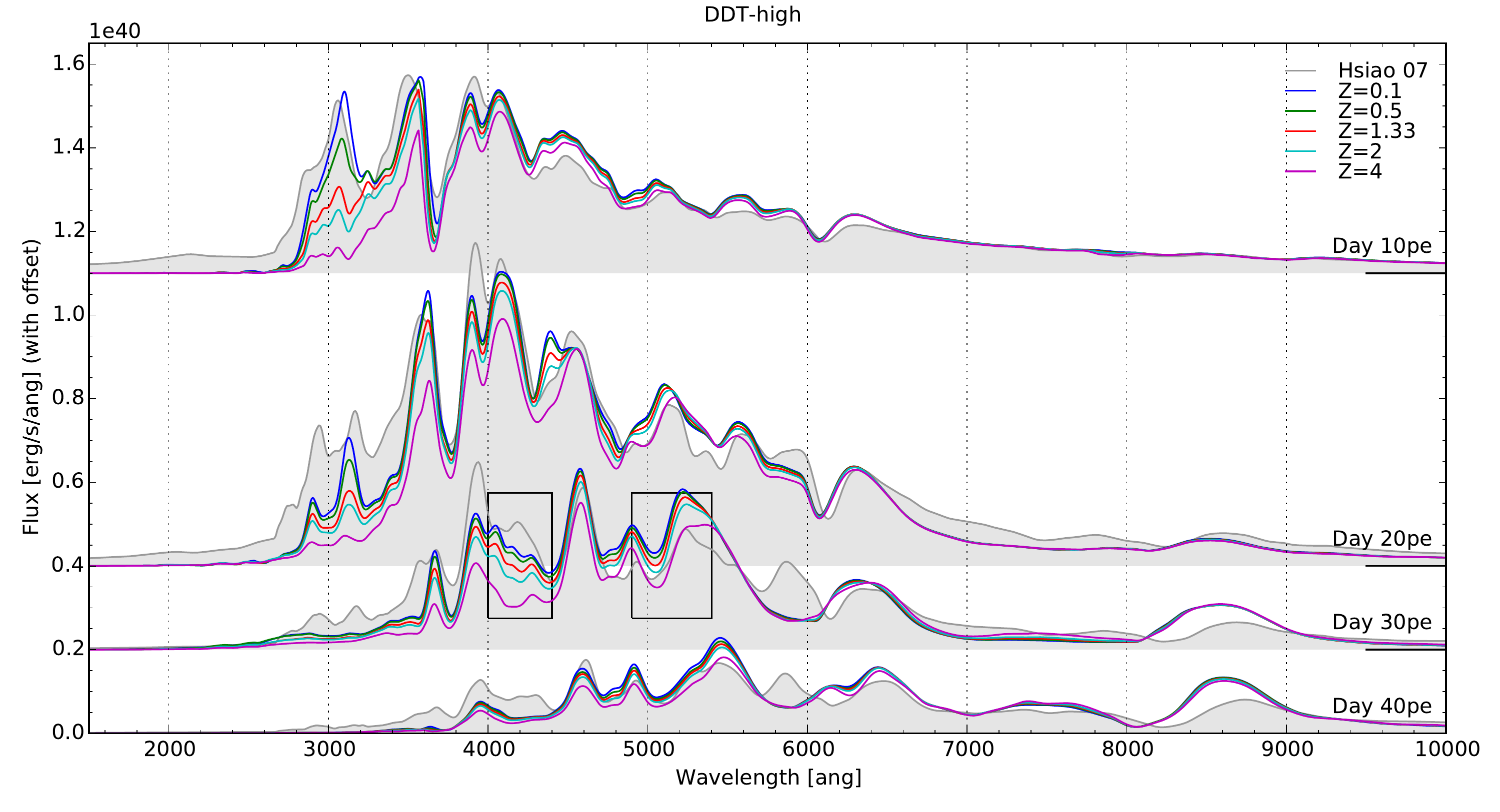}}
\centerline{\includegraphics[width=\textwidth]{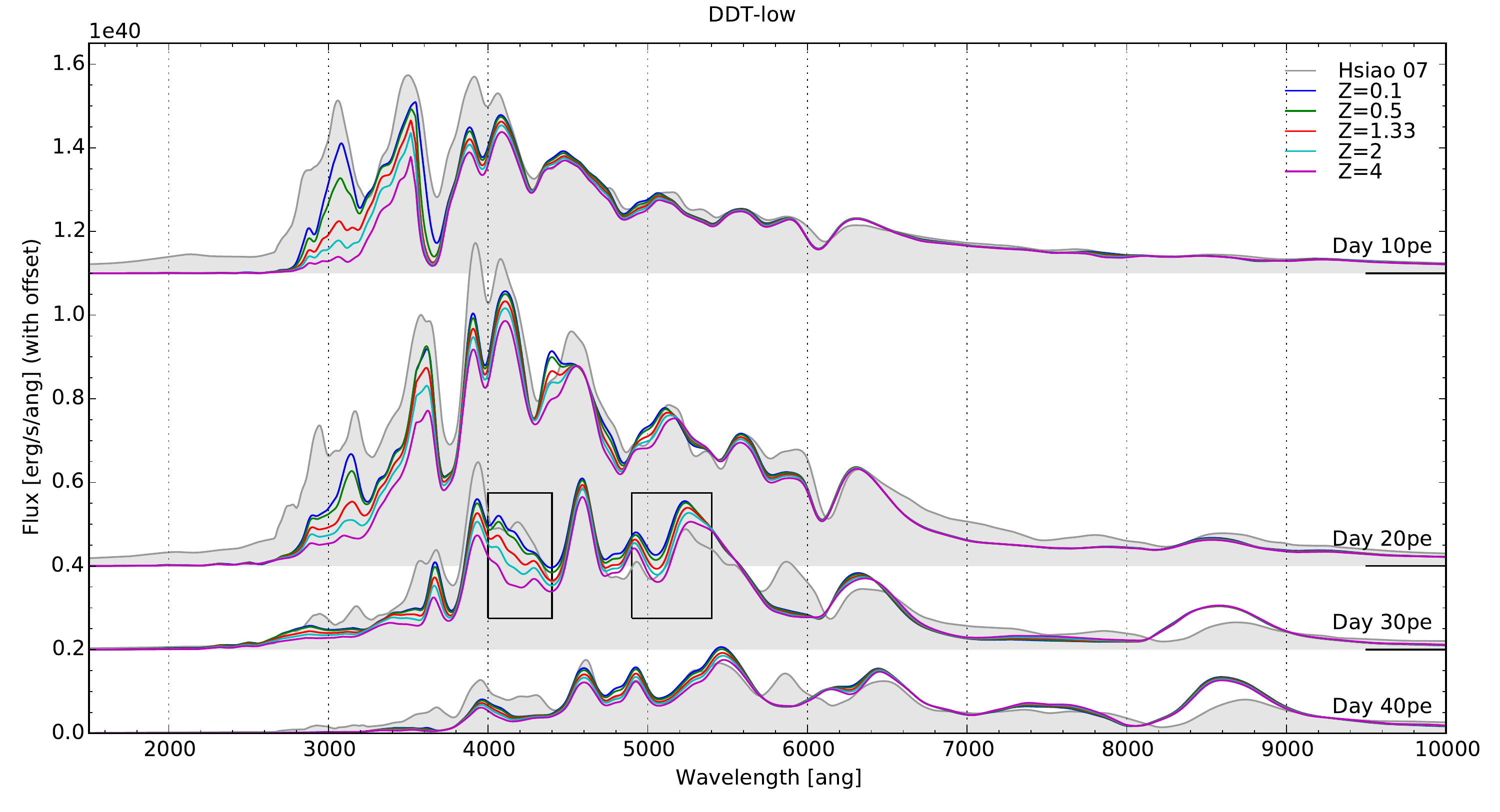}}
\caption{Model spectra of the DDT-high model (\emph{top}) and DDT-low model (\emph{bottom}) for five different progenitor metallicities at day 10, 20, 30, and 40 post explosion (pe).
 For comparison against observational data, the \citet{Hsiao07} spectral templates are plotted in grey.
 The spectral feature in the black box around 4200\,\AA\ and 5200\,\AA\ at day 30\,pe are potential spectral indicators for the progenitor metallicity (see text).
}\label{fig:spectra}
\end{figure*}

\begin{figure*}
\centerline{\includegraphics[width=\textwidth]{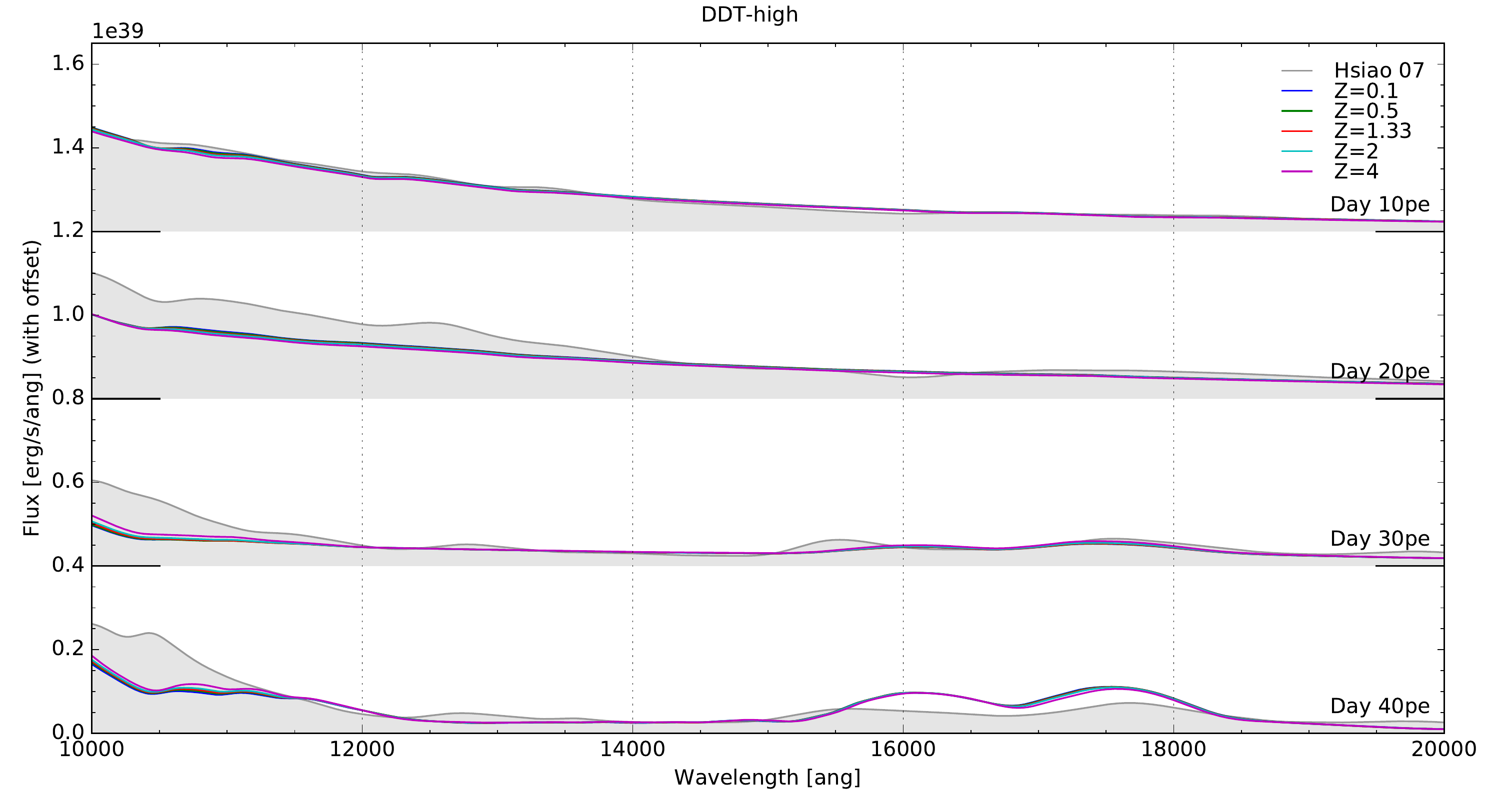}}
\centerline{\includegraphics[width=\textwidth]{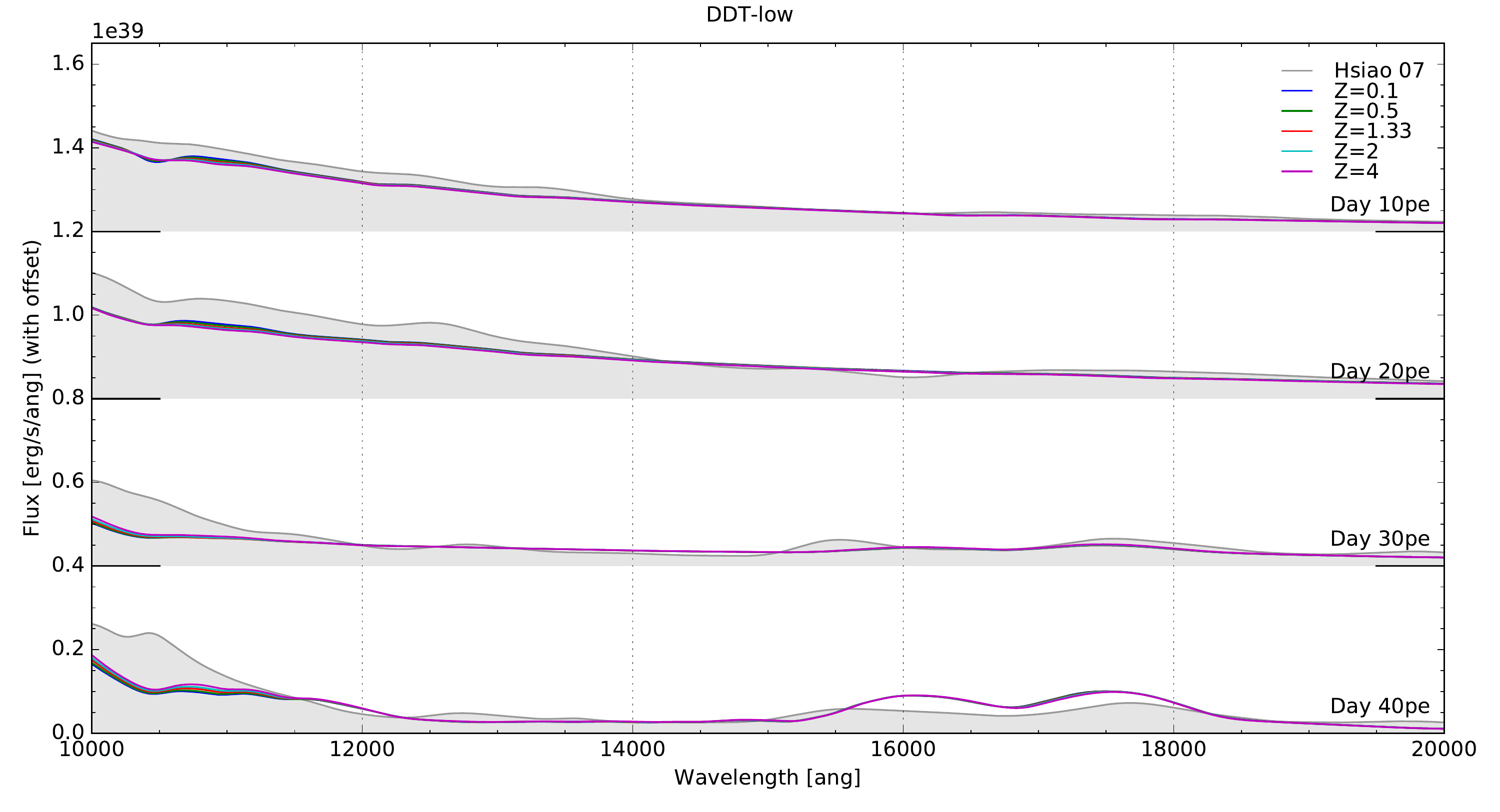}}
\caption{Spectra of Figure~\ref{fig:spectra} extended to the IR range
}\label{fig:spectra-IR}
\end{figure*}

\begin{figure*}
\centerline{\includegraphics[width=\textwidth]{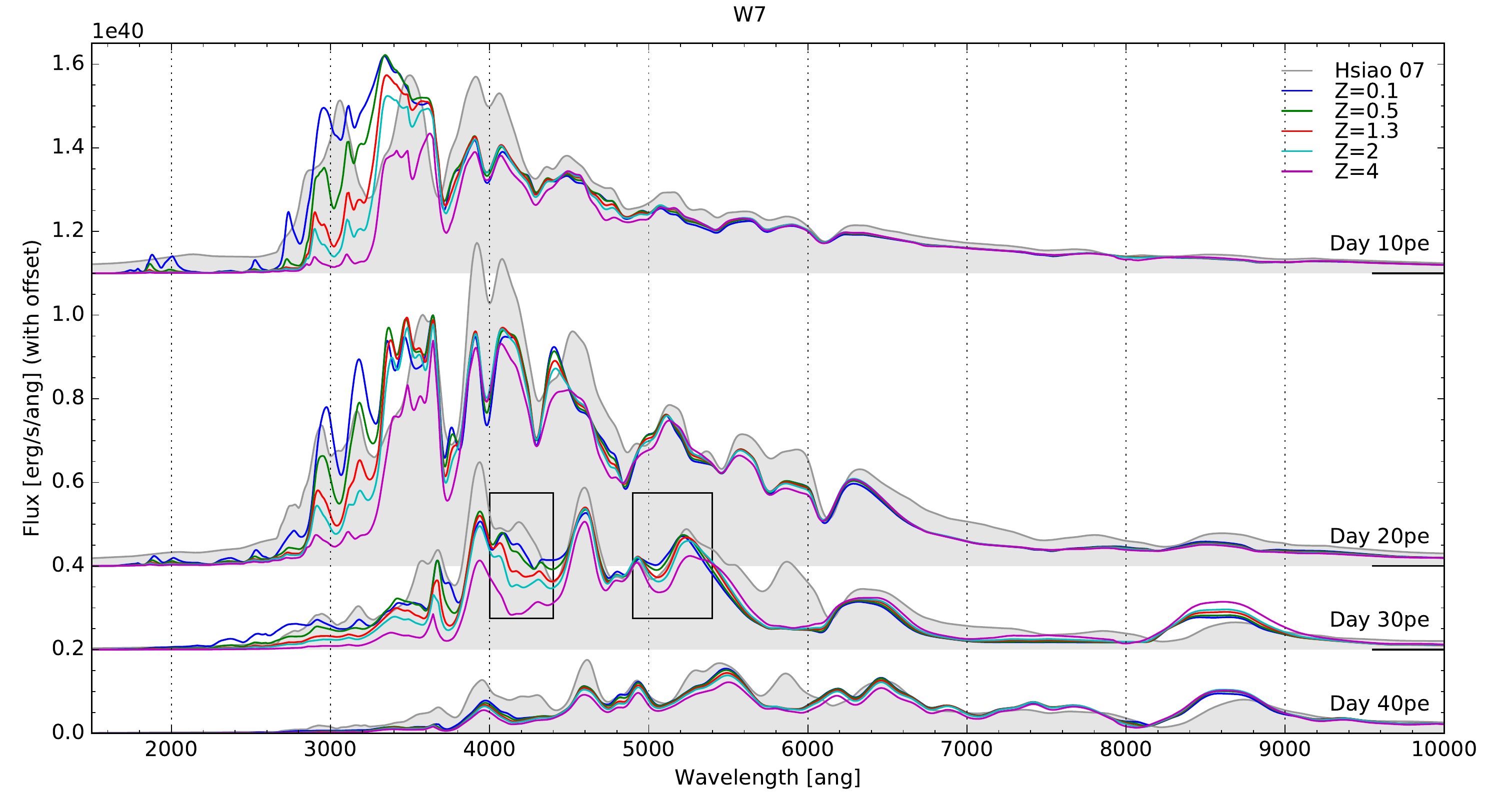}}
\caption{Model spectra of the W7 model for five different pre-explosion metallicities at day 10, 20, 30, and 40 post explosion (pe).
 For comparison against observational data, the \citet{Hsiao07} spectral templates are plotted in grey.
 The spectral features in the black boxes around 4200\,\AA\ and 5200\,\AA\ at day 30\,pe are potential spectral indicators for the progenitor metallicity (see text).
}\label{fig:spectra_w7}
\end{figure*}

\begin{figure*}
\centerline{\includegraphics[width=\textwidth]{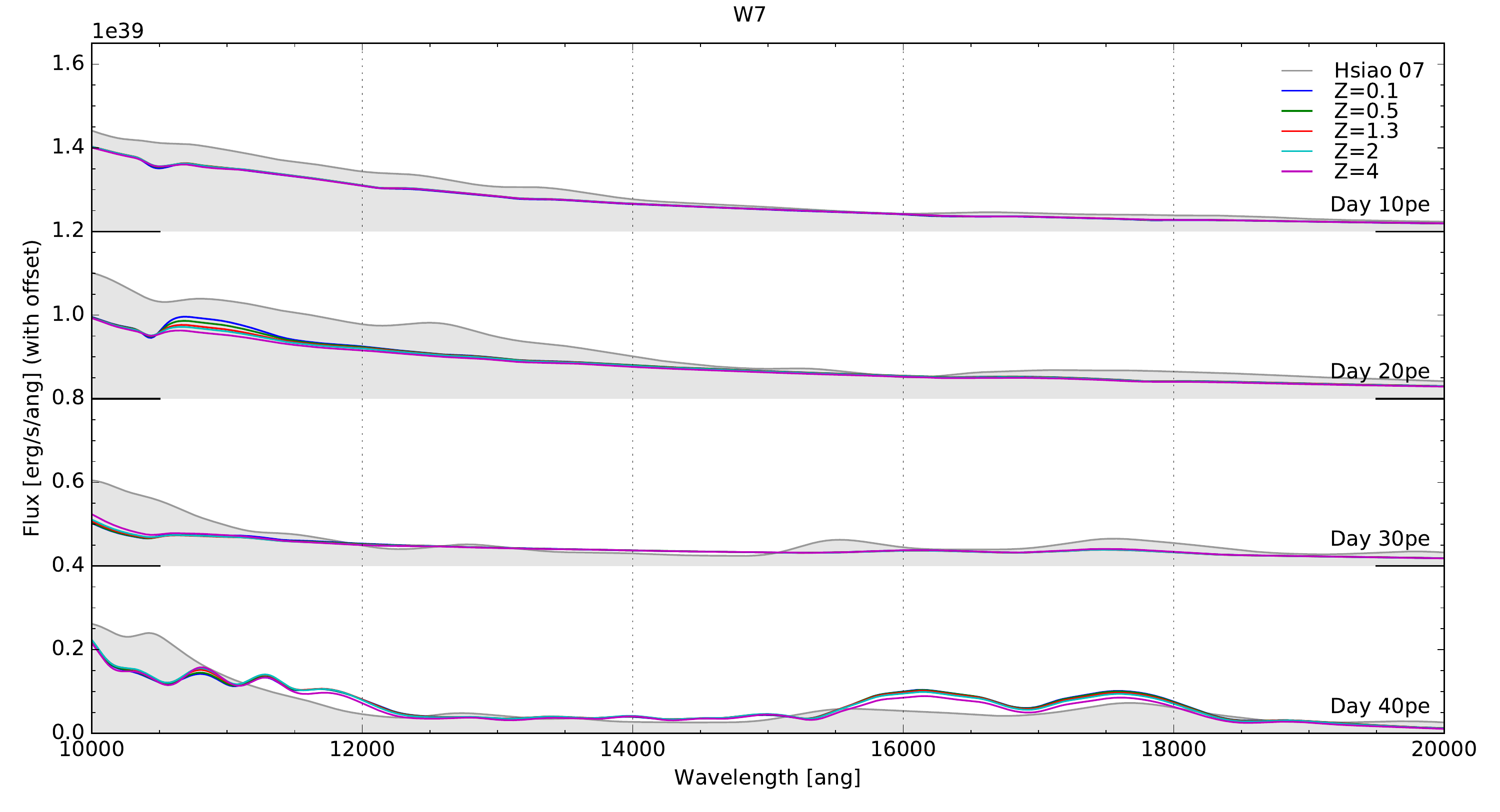}}
\caption{Spectra of Figure~\ref{fig:spectra_w7} extended to the IR range
}\label{fig:spectra_w7-IR}
\end{figure*}

Model spectra, calculated for W7 and the two DDT simulations, with different progenitor metallicities, and at four different epochs post explosion, are shown in Figures~\ref{fig:spectra} through~\ref{fig:spectra_w7-IR}.%

The model spectra are compared against the \cite{Hsiao07} spectral templates.
These templates represent the typical spectral shape of normal \snia, as they are constructed as an average over many observed spectra.
We set day 0 of the Hsiao07 templates (which is the time of maximum brightness) to the time that the B-band luminosity peaks in the model spectra, averaged over the set of progenitor metallicities, and fit a ``distance'' to the imaginary source of the Hsiao07 templates to simultaneously fit the spectra of all epochs.
We do not apply any other corrections, such as stretch or color corrections, to the Hsiao07 templates.

The model spectra generally agree reasonably well with the templates, but there are a few significant discrepancies.
First, none of the model spectra reproduce the far UV flux below 2500\,\AA\ as it is observed on Day 10 and 20 post explosion (pe).
Second, although the Si\,\textsc{ii} feature at 6150\,\AA\ is present in the model spectra none of them accurately reproduces the observed shape of this feature.
It is unclear whether this is a consequence of an under or over produced nucleosynthetic yield or a missing line or combination of lines from the radiative transfer calculation. 
Third, the W7 model spectra do not reproduce the observed strong absorption feature around 3300\,\AA\ at day 10\,pe.

Different progenitor metallicities give rise to variations in the model spectra.
These variations change with time.
Generally, the variations appear to affect the flux more on the blue side of the spectrum than on the red side and, over time, the range in which variations are strong gradually extends further to longer wavelengths.
At day 10\,pe, the DDT and W7 spectra are most effected in the 2600--3600\,\AA\ range.
At days 20 and 30\,pe that window extends to 2600--4500\,\AA\ and 2600--5200\,\AA, respectively.
In the latest spectra, at day 40\,pe, the effect of progenitor metallicity presents only a small amount of variation.

The systematic decrease of blue (short-wavelength) flux with progenitor metallicity is correlated in part to the decline in total \nifs yield with metallicity.
This makes it more difficult to use this region as an indicator of metallicity that is independent of overall \nifs yield.
The total \nifs yield affects the brightness of the light curve as well as the temperature of the radiating ejecta, and thus the spectral color.
At the same time, the progenitor metallicity affects the composition of the \snia ejecta, which may affect the strength of individual spectral features.

\begin{figure*}
\centerline{\includegraphics[width=\textwidth]{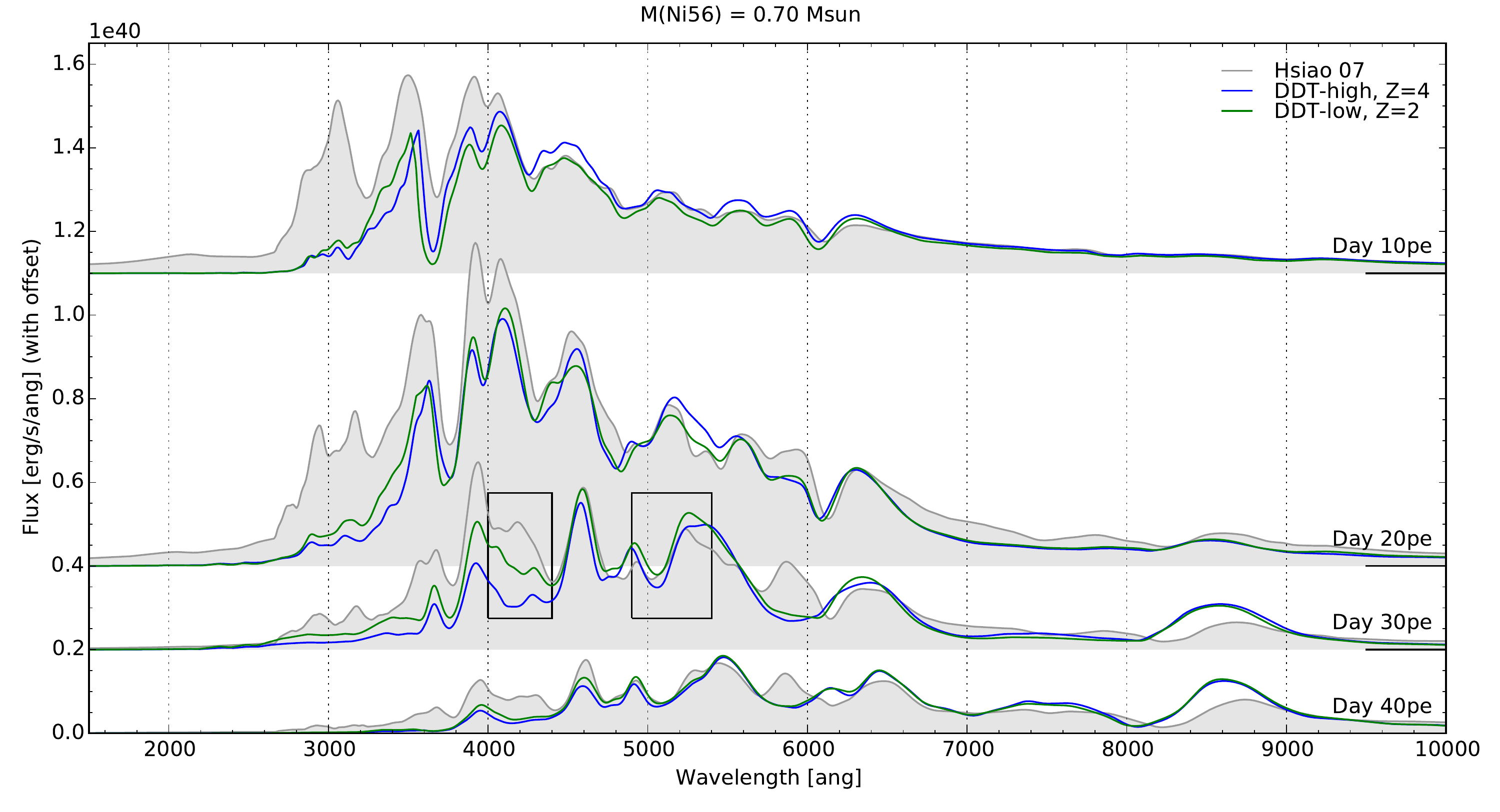}}
\centerline{\includegraphics[width=\textwidth]{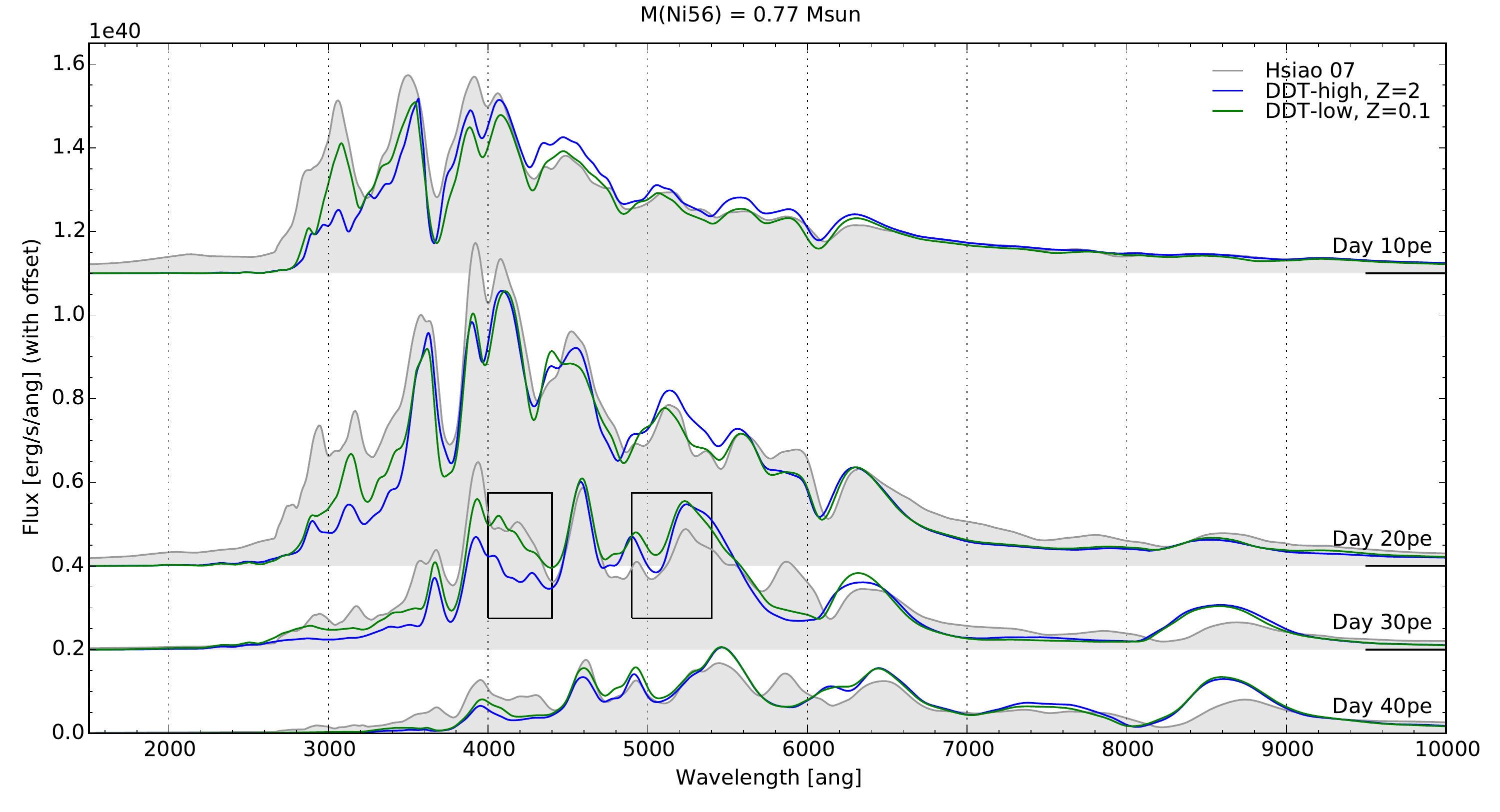}}
\caption{Model spectra at similar \nifs yields, 0.7 $M_\odot$ (\emph{top}) and 0.77 $M_\odot$ (\emph{bottom}), from different DDT models at different progenitor metallicities.
 This comparison, together with Figure~\ref{fig:spectra}, helps to determine which variations in the spectra are due to changes in the abundance profiles of species other than \nifs.
 The most characteristic feature that robustly varies with progenitor metallicity, even at constant \nifs mass, are the absorption features highlighted with black boxes around 4200\,\AA\ and 5200\,\AA\ at day 30\,pe.
}\label{fig:spectra_sameNi}
\end{figure*}

In order to help demonstrate the difference between the \nifs-mass--temperature--color effect and the composition--feature-strength effect, Figure~\ref{fig:spectra_sameNi} compares spectra at similar \nifs yields.
From the top panel spectra at days 10 and 20\,pe, we see that the region of the spectrum around 3000\,\AA\ does not show as strong a metallicity variation in this comparison as it does in Figure~\ref{fig:spectra}.
The case in the bottom panel with slightly higher yield and lower metallicity, shows more metallicity variation in this region.
Together these comparisons show that it will be necessary to carefully control for total \nifs yield in order to use a metallicity indicator from this spectral region.

Features that vary consistently in both Figures~\ref{fig:spectra} and~\ref{fig:spectra_sameNi} are good candidates for robust metallicity indicators that may not require careful control for \nifs yield, which renders the feature useless if the \nifs yield is not known.
Two features that clearly vary with progenitor metallicity at day 30\,pe in Figure~\ref{fig:spectra_sameNi} as well as in all spectra in Figure~\ref{fig:spectra} are 1) an absorption feature around 4200\,\AA, and 2) an absorption feature around 5200\,\AA\@.
These two features are potential \emph{spectral indicators of progenitor metallicity}.
We will examine these features more below.

There are a couple of other features that are promising, but their variation with metallicity is less consistent for both varying (Figure~\ref{fig:spectra}) and fixed (Figure~\ref{fig:spectra_sameNi}) \nifs.
The variation with metallicity around 4400\,\AA\ at day 20\,pe in Figure~\ref{fig:spectra} is much less clear in Figure~\ref{fig:spectra_sameNi}.
The feature around 5500\,\AA\ at day 20\,pe that appears to vary in Figure~\ref{fig:spectra_sameNi} does not vary in the same way in Figure~\ref{fig:spectra}.
For this latter feature it appears that the two DDT models produced different amounts of the chemical elements that are responsible for these features, and so this is not a simple effect of progenitor metallicity.

The Si\,\textsc{ii} P-Cygni feature at 6150\,\AA\ is present in the DDT and W7 model spectra at peak brightness and before.
Even though, as noted before, the observed shape of this feature is not very accurately reproduced in the model spectra, the change in metallicity has very little effect on both the depth of the absorption wing and the height of the emission wing.
This is in agreement with the observation in Figures~\ref{fig:abundance_ratios} and~\ref{fig:total_yields}, namely that the Si abundance remains relatively static by changes in the progenitor metallicity.
This allows for its use as a possible calibration when examining the strengths of other features.

\subsection{Analysis of Spectral Features Around 4200\,\AA\ and 5200\,\AA}

\begin{figure}
\centerline{\includegraphics[width=.5\textwidth]{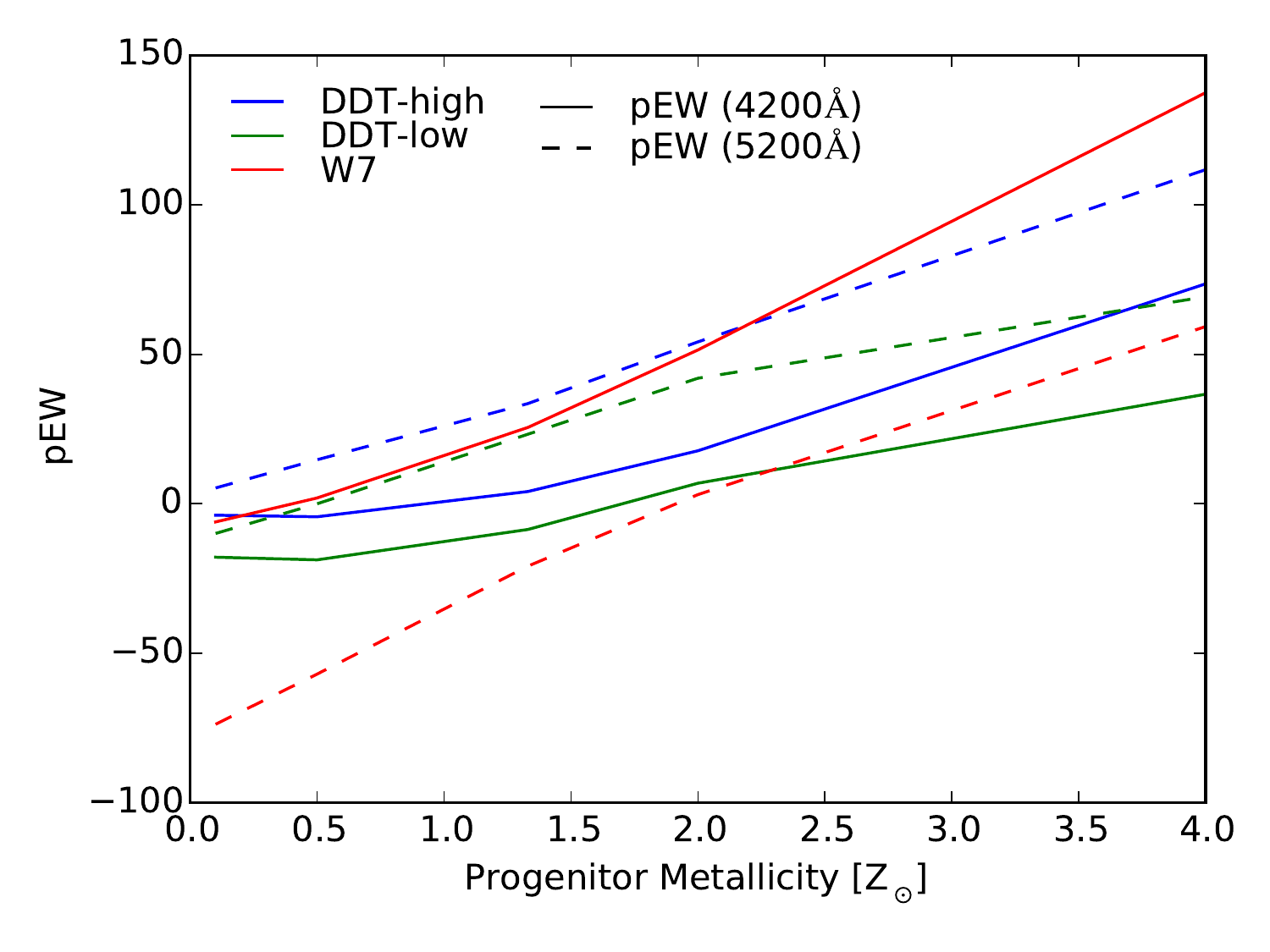}}
\caption{The pseudo equivalent width (pEW) of the spectral features around 4000--4400\,\AA\ (solid lines) and 4900--5400\,\AA\ (dashed lines) as measured in the synthetic spectra of three explosion models DDT-high (blue), DDT-low (green), and W7 (red) at day 30\,pe (see Figures~\ref{fig:spectra} and~\ref{fig:spectra_sameNi}).
 Using the three models investigated in the paper, these two absorption features gradually become stronger with increasing progenitor metallicity, making them good potential candidates for spectral indicators of \snia\ progenitor metallicity.
}\label{fig:pEW}
\end{figure}

Two spectral features were identified above as potential spectral indicators of progenitor metallicity, in all three models: 1) an absorption feature around 4200\,\AA, and 2) an absorption feature around 5200\,\AA; both at day 30\,pe.
Both of these features become systematically stronger with increasing progenitor metallicity.
Figure~\ref{fig:pEW} shows the trend of the feature strength with progenitor metallicity, with the pseudo equivalent width pEW as measure for feature strength.
\begin{equation}
 \textrm{pEW} = \int_{\lambda_a}^{\lambda_b} \frac{F_{\rm pCont}(\lambda) - F(\lambda)}{F(\lambda)} d\lambda ,
\end{equation}
where $F_\textrm{pCont}$ is the pseudo continuum flux level as approximated by a linear interpolation between the flux levels $F$ at beginning and end points of integration $\lambda_a$ and $\lambda_b$.
The interpolation range for the two features is chosen as [4000, 4400] and [4900, 5400] in units of \,\AA\@.

While the dependence of the pEW on metallicity is promising, the variation across metallicities in the $Z/Z_\odot=$0.5 to 2 range is similar to that among the three explosion models considered here.
This means that, in order to use a plot like Figure~\ref{fig:pEW} to determine the metallicity, it is necessary to obtain a control variable that distinguishes among the models.
Unfortunately the most accesible control variable $\Delta m_{15}(B)$ does not appear to fill this role well.
The DDT-high model at $Z/Z_\odot=0.5$ and the DDT-low model at $Z/Z_\odot=2$ have both very similar $\Delta m_{15}(B)$ and very similar 4000-4400\,\AA\ pEWs.
Our limited sample of explosions available here does not allow us to robustly test control parameters.
However, we are hopeful that a larger sample, theoretical or observational, may be able to exploit the trends in pEW shown here to calibrate a measure of metallicity.

\begin{figure*}
\centerline{\includegraphics[width=\textwidth]{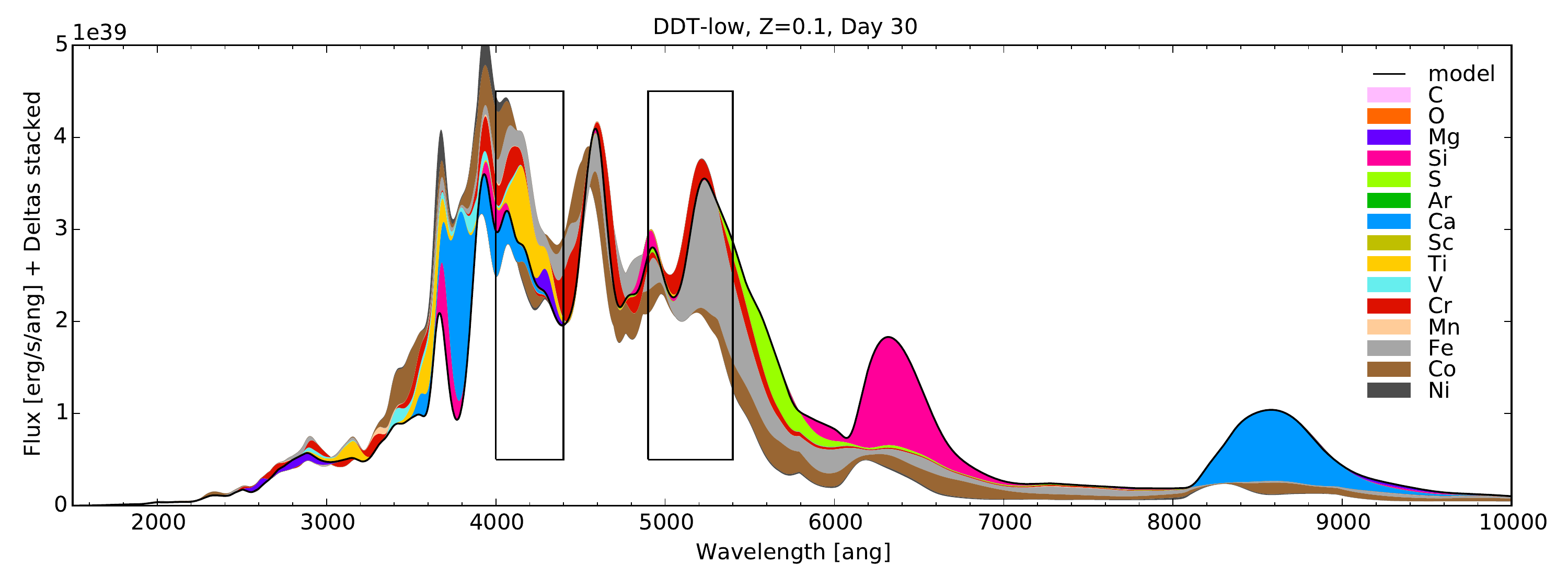}}
\centerline{\includegraphics[width=\textwidth]{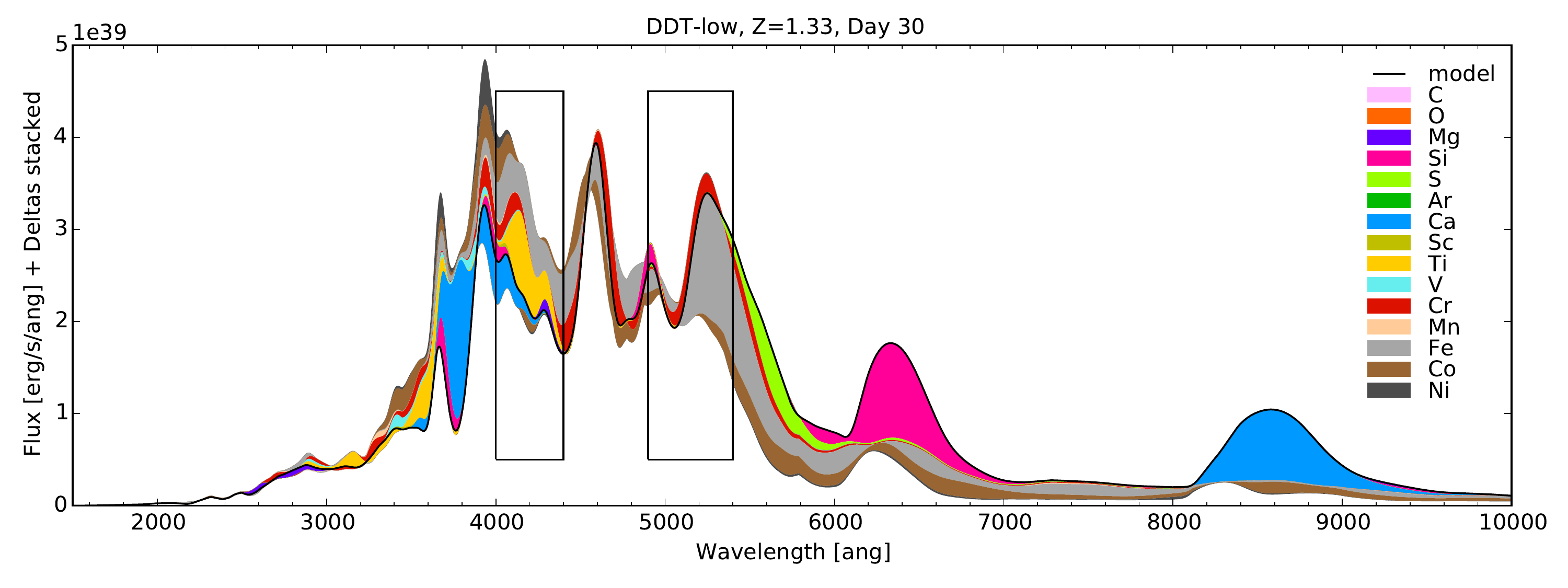}}
\centerline{\includegraphics[width=\textwidth]{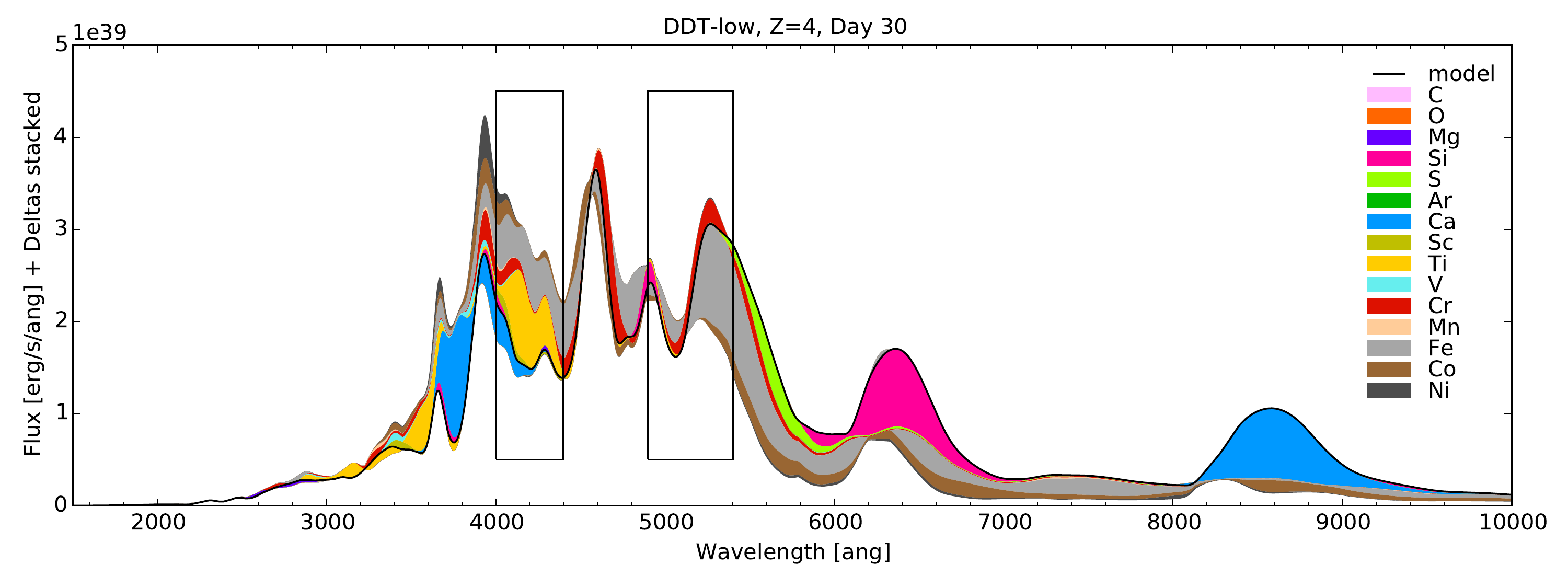}}
\caption{Knock-out spectra for the DDT-low model with metallicity Z/\zsun=0.1 (top), 1.33 (middle), and 4.0 (bottom) at day 30\,pe.
Contributions from line opacity of individual elements to the full spectrum (black solid lines) are represented by colored patches.
These are calculated by removing the line opacities from a single element and recalculating the spectra while keeping the temperatures and population numbers fixed.
Emission contributions are represented by colored patches below the full spectra, and absorption contributions are represented by patches above the full spectra.
The spectral features highlighted with black boxes are two potential spectral indicators for progenitor metallicity.
}\label{fig:knockout_ddt}
\end{figure*}

\begin{figure*}
\centerline{\includegraphics[width=\textwidth]{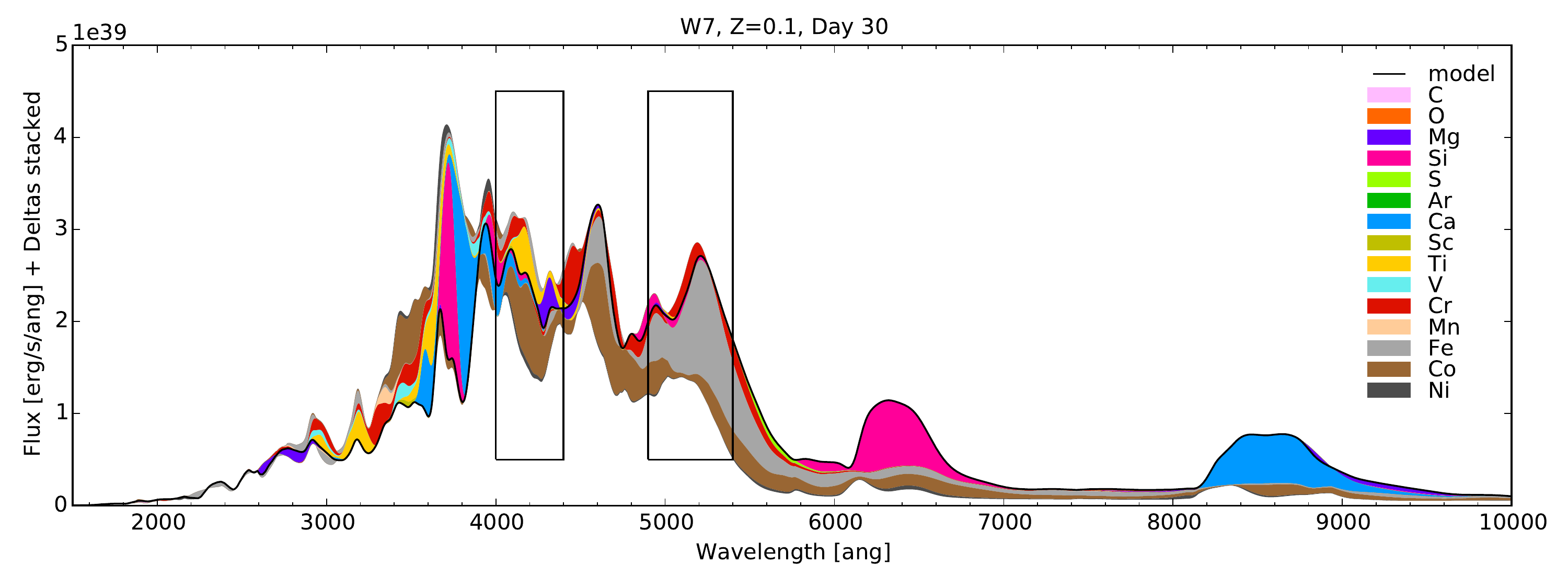}}
\centerline{\includegraphics[width=\textwidth]{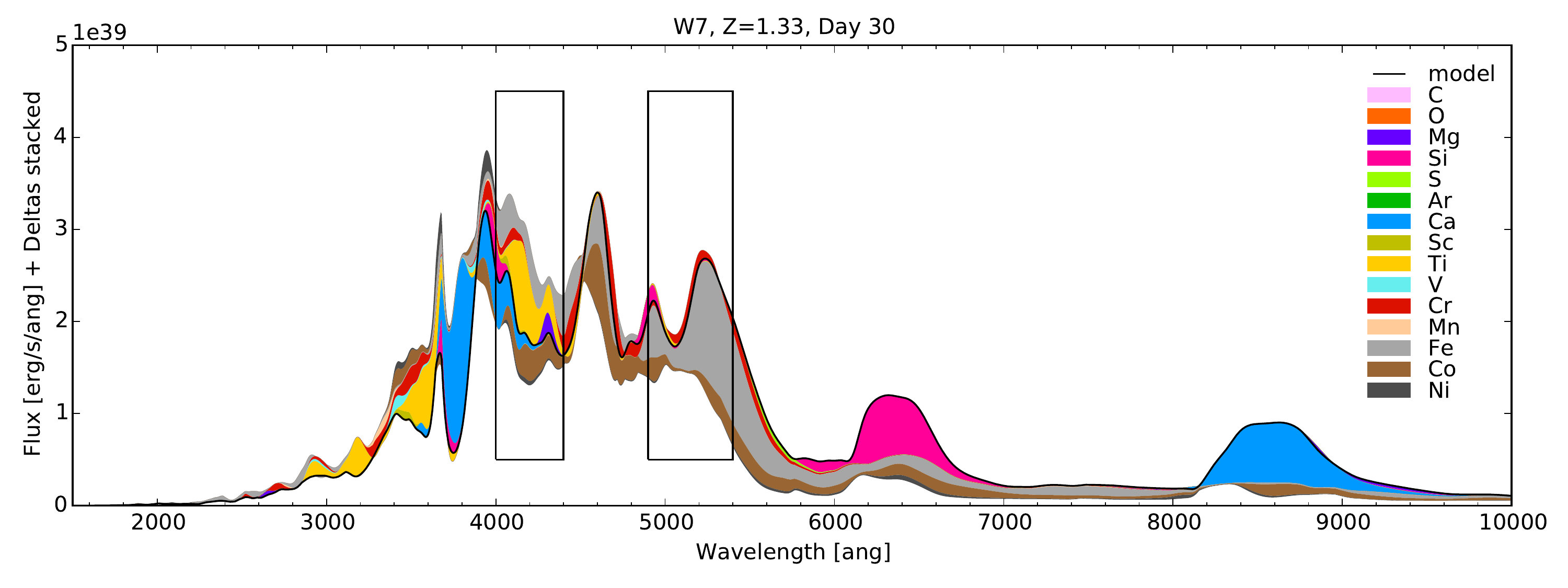}}
\centerline{\includegraphics[width=\textwidth]{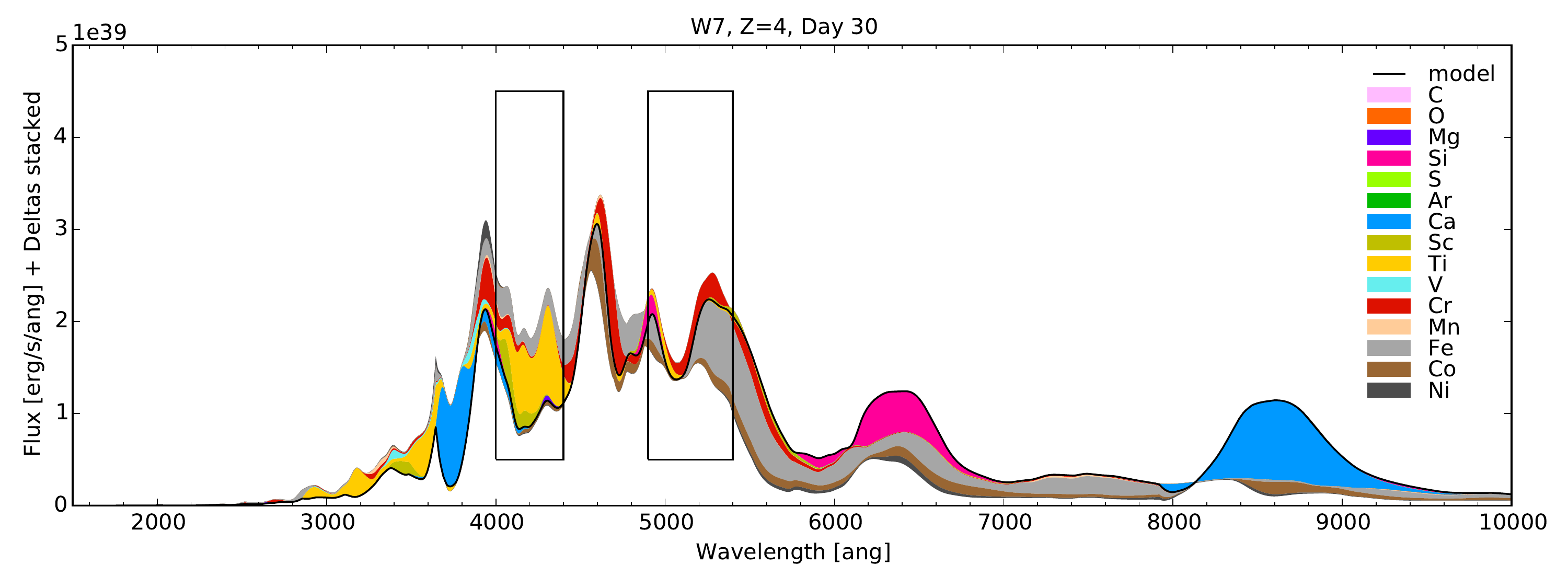}}
\caption{Like Figure~\ref{fig:knockout_ddt} but for the W7 model.
}\label{fig:knockout_w7}
\end{figure*}

In order to investigate which chemical species are responsible for these two specific characteristic features we use knockout spectra (see Section~\ref{sec:radtrans}).
Figures~\ref{fig:knockout_ddt} and~\ref{fig:knockout_w7} displays a series of knockout spectra for the DDT-low and the W7 model at day 30\,pe for Z/\zsun = 0.1, 1.33 and 4.

These knockout spectra are created by removing the opacity from all transition lines of a single chemical element from the radiative transfer calculation and repeating the simulation with fixed temperatures, electron densities, and population numbers as in the original calculation.
The recalculated spectra show the effect of the absence of opacity.
This allows us to identify what and how strongly different elements are involved in forming specific spectral features.

In the knockout spectra, Figures~\ref{fig:knockout_ddt} and~\ref{fig:knockout_w7}, the full spectra are represented by the solid black line.
The color patches represent the knockout effect that each element has on the full spectrum.
Patches below the spectra indicate emission features: the flux level is that much lower in the absence of opacity from that species.
In the same way, patches above the full spectrum represent absorption features. 

Using the knockout spectra, we see that the two identified spectral indicators of progenitor metallicity, around 4200\,\AA\ and 5200\,\AA\ at day 30\,pe, can be associated with Titanium and Iron production, respectively.
The higher the progenitor metallicity the stronger the Titanium and Iron abosption features appear in the spectra.
Iron also shows a shift from emission to absorption in this feature as metallicity increases.
Looking back at Figures~\ref{fig:abundance_profiles} and~\ref{fig:ti_profiles}, we see that indeed the Ti and Fe abundances increase with the progrenitor metallicity in the outer regions of the ejecta, i.e., at velocities $ > 10^9$ cm/s.

\subsection{Spectra Ratios}

\cite{Foley_2013}, and more recently, \cite{Graham15} investigated the two \sneia\ SN 2011fe and SN 2011by.
These two objects exhibited remarkably similar optical spectra and decline times, and even peak brightnesses after accounting for possible errors in the measured distance to 2011by.
Thus the two objects become nearly identical.
However, the two objects display differences in the near UV regions of their spectra. 
\cite{Graham15} propose that one possible source of these differences could be a difference in progenitor metallicity. 
To investigate this proposal, they first normalize the two spectra across 4000--5500\,\AA\ and then take the ratio of the two spectra, 11fe:11by.
This ratio is then compared to ratios of model spectra at difference metallicites produced by \cite{Lentz2000}.
By comparison, they find the best match between the observed spectral ratio and the model spectra corresponds to a factor of 30 difference in progenitor metallicity \citep[Figure 8]{Graham15}. 
In this section, we discuss how well our models reproduce the observed UV spectra in these cases, and challenges in making this type of comparison when metallicity and explosion strength (\nifs\ mass) are allowed to vary independently.

\begin{figure*}
\centerline{\includegraphics[width=.5\textwidth]{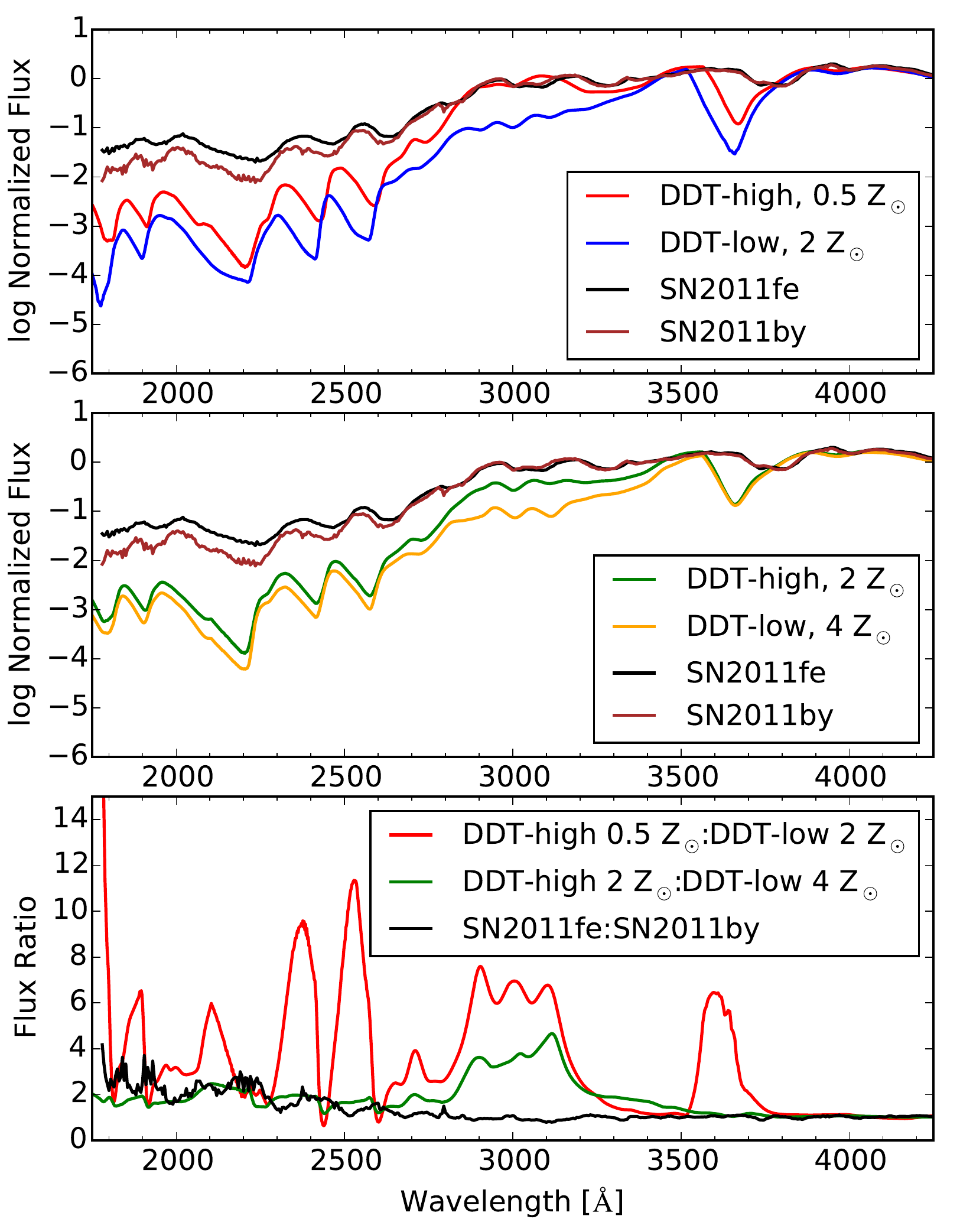}
			\includegraphics[width=.5\textwidth]{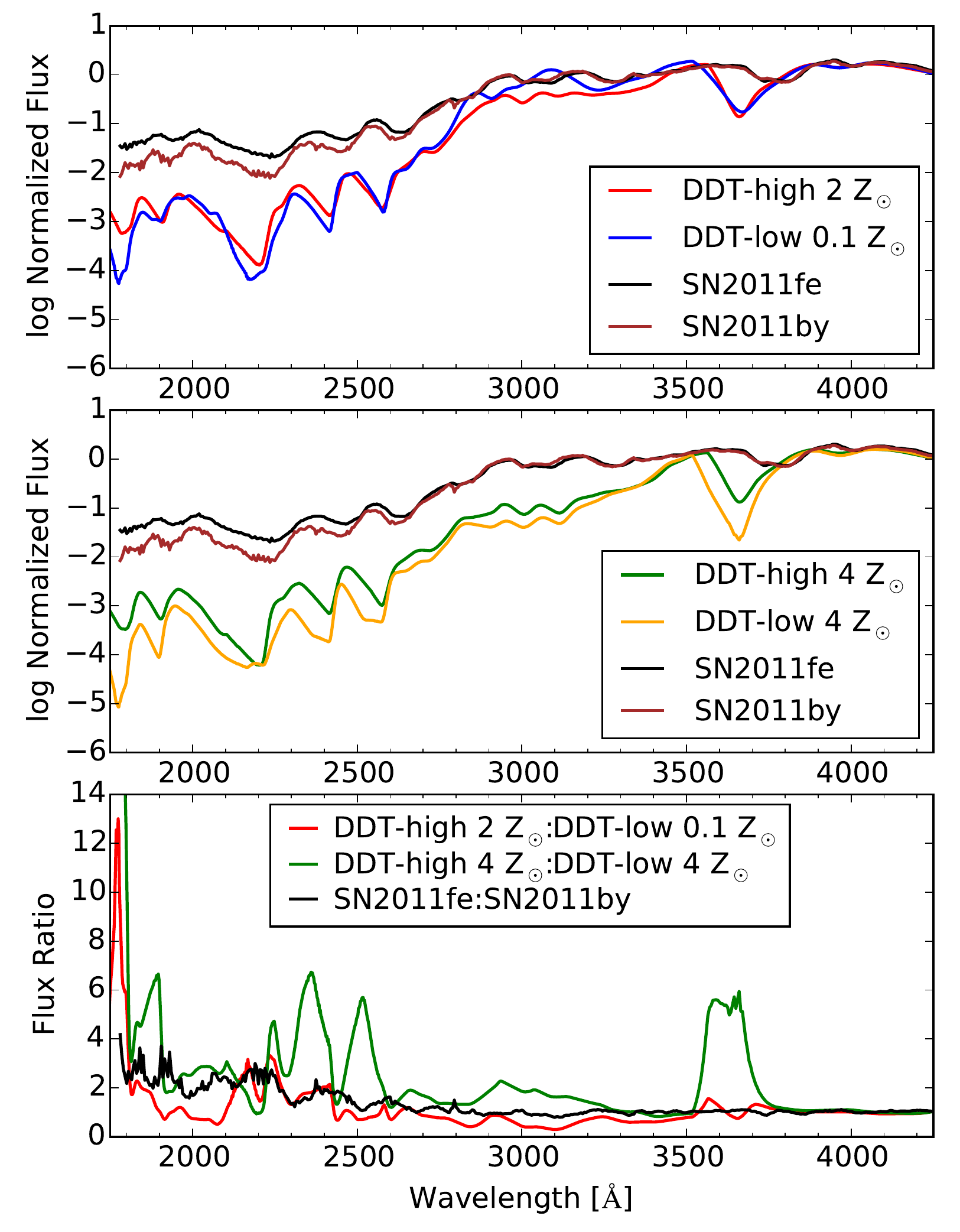}}
\caption{\emph{Top and middle}:
The normalized maximum light NUV spectra of pairs of spectra chosen for their similar decline times (left) and similar optical spectral features at maximum brightness (right).
The spectra of SN2011fe and SN2011by are also shown for comparison.
\emph{Bottom}:
Ratios of the maximum light NUV spectra of each pair chosen for similar decline times (left) and similar optical spectral features (right) with the ratio of SN2011fe to SN2011by included for comparison.
Ratios are shown for the top panel pair (red) and middle panel pair (green).}
\label{fig:spectra_ratio_decline}
\end{figure*}

In Figure~\ref{fig:spectra_ratio_decline}, we take 4 pairs of spectra from our models at day 20\,pe and follow a similar procedure as \cite{Graham15}. 
Two pairs, DDT-high Z = 0.5 Z$_{\odot}$ \& \ DDT-low Z = 2.0 Z$_\odot$ and DDT-high Z = 2.0 Z$_\odot$ \& \ DDT-low Z = 4.0 Z$_\odot$, have similar decline times as shown in Figure~\ref{fig:phillips}.
The other two pairs, DDT-high Z = 2.0 Z$_\odot$ \& \ DDT-low Z = 0.1 Z$_\odot$ and DDT-high Z = 4.0 Z$_\odot$ \& \  DDT-low Z = 4.0 Z$_\odot$, have similar optical spectral features at day 20\,pe. 
Each of the spectra are normalized across the 4000 -- 5500\,\AA\ region.
The top and middle panels of the left column of Figure~\ref{fig:spectra_ratio_decline} show the normalized spectra in log space of each pair selected for similar decline times.
The top and middle panels of the right column of Figure~\ref{fig:spectra_ratio_decline} show the normalized spectra in log space of each pair selected for similar optical features.
Also included in the top and middle panels of both columns of Figure~\ref{fig:spectra_ratio_decline} are the spectra of SN2011fe at -2 days from peak brightness \citep{Mazzali14} and SN2011by at -1 day (both Melissa Graham, priv.\ comm.).
These  spectra were first binned in increments of 5 \,\AA\ and then normalized in the same way as our calculated spectra.
Our calculated spectra share a similar sequence of features, especially blueward of about 3000\,\AA, with these two observed spectra.
However, the features in the model spectra are both stronger and located at slightly shorter wavelengths.
Also the overall relative flux level blueward of about 2700\,\AA\ is about a factor of 10 lower in the model spectra.

The bottom panels of both columns in Figure~\ref{fig:spectra_ratio_decline} show the ratios taken between each pair as well as the ratio of SN2011fe to SN2011by.
One of the most prevalent differences in the model spectra is in the 3000\,\AA\ region, which is one of the regions most strongly effected by varying metallicity.
The model pairs selected to have the same decline time show a factor of 4-6 ratio in this region, while the observed pair shows no significant difference.
Comparing the ratios of pairs, DDT-high Z = 0.5 Z$_{\odot}$ \& \ DDT-low Z = 2.0 Z$_\odot$ (Figure~\ref{fig:spectra_ratio_decline}, bottom left panel, red curve) and DDT-high Z = 4.0 Z$_\odot$ \& \  DDT-low Z = 4.0 Z$_\odot$ (Figure~\ref{fig:spectra_ratio_decline}, bottom right panel, green curve), demonstrates the difficulty in separating metallicity from \nifs yield variations.
The location of the peaks in these two pairwise comparisons is similar, having ratio peaks at about 1900, 2350, 2500, and 3600\,\AA\@.
However, while the first pair are at different metallicities, the second pair are at the same metallicity but with different \nifs\ yields.
A possible factor in the similarity can be seen in the $^{56}$Ni yields.
Looking at Table~\ref{tab:yields_grouped}, it is clear that none of the four cases produced equal amounts of $^{56}$Ni, though the ratios of $^{56}$Ni produced in each pair are very similar at~$\approx$~1.1.
Another contribution comes from the apparent blueshifting of features in the DDT-low spectra compared to the same features in the DDT-high spectra.
In Section~\ref{sec:spectra}, we have discussed that many of the changes in bluer side of the NUV region are caused by temperature effects from the changing $^{56}$Ni yield.
Since changes in spectral features from this temperature effect occur alongside changes from the varying ejecta composition, isolating the two effects by taking spectral ratios appears challenging.
A more conclusive comparison will require model spectra that are more similar to the observations in this spectral range.


\section{Conclusions and Discussion}

We have utilized a combination of multidimensional hydrodynamics calculations, nucleosynthetic particle post-processing calculations, and radiative transport calculations in a preliminary investigation into the role that progenitor metallicity plays in the chemical composition of \snia ejecta, and the consequences that has for their light curves and spectra.
These calculations were performed on two 2-dimensional DDT models, as well as on the 1-dimensional W7 model.
Lagrangian tracer particles were produced for the two DDT models using the FLASH hydrodynamics software instrument.
These tracer particles as well as the W7 temperature-density histories were then post-processed using TORCH with a 225 member nuclear reaction network to calculate nucleosynthetic yields and ejecta abundance profiles.
We have used the PHOENIX radiation transport software to compute light curves and spectra for these ejecta profiles.
With this approach we examine the effects of progenitor metallicity on the nucleosynthetic yields of the explosion and on the synthetic light curves and spectra.

Our approach is different from past theoretical studies that have focussed on the effects of metal content in the ejecta on synthetic spectra (see \citet{Brown15}, and references therein) in two ways.
First, we calculate self-consistent nucleosynthetic yields from explosion simulations by changing the progenitor metallicity instead of making modifications to post-explosion abundance profiles.
This takes into account any effects of progenitor metallicity on the \nifs production, which affects post explosion ejecta temperatures, and therefore the shape and color of synthetic light curves and spectra, and in particular in the UV\@.
Second, instead of focussing on the UV flux, we explore the effect of progenitor metallicity on the strength of spectra features in the optical.
The UV flux has been found to be sensitive not only to metal content in the ejecta, but also to other properties like the density structure and the type of explosion model \citep[and references therein]{Brown15}, a finding that is confirmed also in this paper.

We find that two aspects of SN Ia spectra present challenges to finding spectral indicators of progenitor metallicity or neutron excess.
The first is that SN Ia spectra are formed by a large number of blended lines from a variety of species.
This makes isolating individual features challenging, more so in some spectral ranges than others.
The second challenge is that while the abundances of both the unburned material and the IME material vary systematically with metallicity, as expected, the resulting variation in spectral features is confused with the variation
of the total $^{56}$Ni yield with metallicity.
Our inclusion of explosions with different yields allows us to explore this ambiguity.
Higher metallicity, which provides a larger neutron excess \citep{timmes_2003_aa}, leads to a lower yield of $^{56}$Ni, instead favoring stable Fe-group yields.
This reduction in the energy source for the photospheric phase causes a lowering of the ejecta temperature, and reduces the overall flux in the UV\@.
Isolating individual features, which vary due to changes in yields with metallicity, from this overall shift in flux intensity is challenging, especially in the early-time UV spectrum.

Despite these challenges, we find two spectral features that show promise as potential robust spectral indicators of progenitor metallicity.
These features stand out as metallicity indicators even when comparing explosions with similar total $^{56}$Ni, a comparison for which many other features are the same (see Figure~\ref{fig:spectra_sameNi}).
The most promising feature we have found is a Ti feature at day 30 post explosion that extends from 4000-4400\,\AA, based on the knock-out spectrum.
The pEW for this feature (Figure~\ref{fig:pEW}) shows a robust and mostly consistent variation with metallicity for each of the three explosion simulations studied here.
However, a control for overall yield will still be necessary for this to be used as a precision indicator of metallicity.
The small sample of explosion models computed here does not allow evaluation of a suitable control variable, but a broader set of computed cases or an observational collection of spectra may.

Our 2D DDT models of SNe Ia produce spectra that are similar to those of observed SNe Ia in many respects (Figure~\ref{fig:spectra}), in some regions better than the spectrum of W7 (Figure~\ref{fig:spectra_w7}).
It is important to note that we are not fitting spectra, so that a precision match is not the goal, rather similar overall spectral features.
One of the main differences between the 2D DDT and W7 is apparent in Figure~\ref{fig:abundance_profiles}, where the DDT models have intermediate mass and iron-group material extending out to higher velocities than W7.
This appears to make the absorption and emission features in the 2D DDT spectrum, especially in the blue portion of the spectrum, more similar to those in observed SNe~Ia.

There are several trends in yields that result from changing progenitor metallicity. 
Table~\ref{tab:yields_grouped} shows that for all three models, increasing metallicity causes a decline in the production of intermediate mass elements and an increase in the production of iron-group material.
However, as is expected with the increased neutronization, the increased production of these stable iron-group elements comes at the cost of reduced production of $^{56}$Ni.
Also at higher metallicities, the iron-group material produced in incomplete Si burning extends further out into the IME layers of the ejecta.
The combination of the increased opacity from the iron-group material and the reduced production of $^{56}$Ni results in light curves that are both slower rising and dimmer than those of lower metallicities.
However, we find only an $\sim .10$ mag dependence of the peak luminosity between the lowest and highest progenitor metallicities, and with our small sample size we are hesitant to make strong comparisons with observations such as \cite{Kelly_2010}. 
Although the overall yield of intermediate mass elements declines with increasing metallicity, the yield and distribution of Si remain relatively static.
Consequently, the \snia characteristic Si P-Cygni feature at 6150\,\AA\ remains mostly unchanged with metallicity.

\acknowledgements
This work is supported in part at the University of Chicago by the National Science Foundation under grant AST-0909132, and under grant PHY-0822648 for the Physics Frontier Center ``Joint Institute for Nuclear Astrophysics'' (JINA).
ACC acknowledges support from the Department of Energy under grant DE-FG02-87ER40317.
Some of the software used in this work was in part developed by the DOE-supported ASC/Alliances Center for Astrophysical Thermonuclear Flashes at the University of Chicago.
We thank Nathan Hearn for making his QuickFlash analysis tools publicly available at http://quickflash.sourceforge.net.

\bibliography{master,timmes_master,townsley_master,vanrossum}

\end{document}